\documentclass[onecolumntraditional]{aa}

\usepackage{amsmath}
\usepackage{graphicx}
\usepackage{natbib}
\usepackage{txfonts}
\usepackage{indentfirst}

\newcommand{\eq}[1]{\begin{equation}  #1 \end{equation}}
\newcommand{\eqs}[1]{\begin{equation} \begin{split} #1 \end{split} \end{equation}}

\newcommand{\br}[1]{\left( #1 \right)}

\newcommand{\bb}[1]{\left[ #1 \right]}
\newcommand{\ba}[1]{\left\langle #1 \right\rangle}
\newcommand{\bs}[1]{\left| #1 \right|}
\newcommand{\dd}{{\rm d}}
\newcommand{\expo}[1]{~{\rm e}^{ #1 }}
\newcommand{\vek}[1]{\mbox{\boldmath $#1$}}
\newcommand{\svek}[1]{\mbox{\boldmath \scriptsize $#1$}}  
\newcommand{\ic}{{\rm i}}

\newcommand{\vz}{{\vek{z}}}
\newcommand{\vzs}{{\vek{z}^*}}
\newcommand{\vv}{{\vek{v}}}
\newcommand{\vvs}{{\vek{v}^*}}
\newcommand{\va}{{\vek{a}}}
\newcommand{\vas}{{\vek{a}^*}}
\newcommand{\vb}{{\vek{b}}}
\newcommand{\vbs}{{\vek{b}^*}}

\begin{document}



\title{Relations between three-point configuration space shear and convergence statistics}

\author{X. Shi \inst{1,2} \and P. Schneider \inst{1} \and B. Joachimi \inst{3}}

\offprints{X. Shi,\\
    \email{xun@astro.uni-bonn.de}
}

\institute{Argelander-Institut f\"ur Astronomie (AIfA), Universit\"at Bonn, Auf dem H\"ugel 71, 53121 Bonn, Germany  
\and International Max Planck Research School (IMPRS) for Astronomy and Astrophysics at the Universities of Bonn and Cologne
\and Institute for Astronomy, University of Edinburgh, Royal Observatory, Blackford Hill, Edinburgh, EH9 3HJ, U.K. }

\date{Received ?? / Accepted ??}

\abstract{
  With the growing interest in and ability of using weak lensing studies to probe the non-Gaussian properties of the matter density field, there is an increasing need for the study of suitable statistical measures, e.g. shear three-point statistics. In this paper we establish the relations between the three-point configuration space shear and convergence statistics, which are an important missing link between different weak lensing three-point statistics and provide an alternative way of relating observation and theory. The method we use also allows us to derive the relations between other two- and three-point correlation functions. We show the consistency of the relations obtained with already established results and demonstrate how they can be evaluated numerically. As a direct application, we use these relations to formulate the condition for E/B-mode decomposition of lensing three-point statistics, which is the basis for constructing new three-point statistics which allow for exact E/B-mode separation. Our work applies also to other two-dimensional polarization fields such as that of the Cosmic Microwave Background.
}

\maketitle

\section{Introduction}
\label{sec:introduction}

The statistical study of the weak gravitational lensing effect by the large-scale structure, also called cosmic shear, is one of the established tools to investigate the matter distribution of the Universe and to constrain the parameters of cosmological models. The undergoing boosting of weak lensing survey size and quality gradually enables one to explore the small, non-linear scales where a wealth of signal lies \citep{ber02,pen03,jarvis04,semboloni11}. Along with this trend, more and more effort has been put into the development of statistical measures that can probe the non-Gaussian signal arising from the non-linear growth of structure on those scales. Among them, the three-point (3-pt) statistics may be the most widely used. 

Two basic quantities considered in gravitational lens theory are the convergence $\kappa$ and the shear $\gamma$. Defined as the dimensionless surface mass density, $\kappa$ is a weighted projection of the three-dimensional (3D) matter density contrast $\delta$. The shear $\gamma$, on the other hand, is directly accessible from observations. Therefore, the theoretical framework of gravitational lensing should include the relation between configuration space $\kappa$ and $\gamma$ statistics as well as the one relating configuration space statistics to their Fourier space counterparts. At the level of two-point (2-pt) statistics, such relations have already been established. For 3-pt statistics, the relation between the shear 3-pt correlation functions ($\gamma$3PCFs) and the convergence bispectrum, which is the Fourier counterpart of the 3-pt convergence correlation function ($\kappa$3PCF), has been derived by \citet{s05}. The other non-trivial relation, the one between $\gamma$3PCFs and $\kappa$3PCFs, is still missing. One purpose of this work is to establish this missing link.    

How to perform E/B-mode decomposition is also a major concern of the weak lensing community. For observational data an E/B-mode decomposition provides a necessary check on the possible systematics \citep[e.g.][]{crittenden02,pen02}. In recent years there have been several efforts to construct better statistics which allow for an E/B-mode decomposition at the 2-pt level \citep{SK07,eifler10,fu10,s10}. They all use weight functions to filter the shear 2-pt correlation functions ($\gamma$2PCFs), and the condition for E/B-mode decomposition transforms to a condition on the weight functions. Such a condition at the 3-pt level is also missing so far. We will see that with the aid of the relation between the $\gamma$3PCFs and the $\kappa$3PCFs, one can easily formulate this condition.

The paper is organized as follows: In Sect.$\,$2 we show how the relation between the $\gamma$3PCF and the $\kappa$3PCF is obtained. In Sect.$\,$3 we investigate the correspondence between the derived relation and already established results. We then extend our results to other $\gamma$3PCFs in Sect.$\,$4, and in Sect.$\,$5 we present an application of the 3-pt relations, deriving the condition for E/B-mode separation of 3-pt shear statistics. How these relations can be numerically evaluated is demonstrated in Sect.$\,$6, and we conclude in Sect.$\,$7.

\section{Relation between 3-pt $\gamma$ and $\kappa$ correlation functions}
\label{sec:relation}
\subsection{The form of the relation}
At the 2-pt level, the relation between the configuration space shear and convergence statistics is the $\xi_+-\xi_-$ relation \citep{crittenden02,s02} 
\eq{
\label{eq:xi+xi-}
\xi_- (x) =  \int \dd y\; y\; \xi_+ (y)\; \bb{\frac{4 x^2-12 y^2}{x^4} H(x-y) + \frac{\delta^{(1)}_{\rm D}(x-y)}{x} } \,,
}
where $H$ and $\delta^{(1)}_{\rm D}$ are Heaviside function and 1D Dirac delta function, respectively. The functions $\xi_+$ and $\xi_-$ are defined as $\xi_+ (x) := \langle \kappa\kappa \rangle (|\vek{x}|)$ and  $\xi_-(x) := \langle \gamma\gamma\rangle (\vek{x}) \expo{-4\ic\phi_x}$, with $\phi_x$ being the polar angle of the separation vector $\vek{x}$, and $\ba{\ }$ indicating the ensemble average. Note that the shear $\gamma$ is a spin-2 quantity, i.e. it gets multiplied by a phase factor $\expo{-2\ic\phi_x}$ when the coordinate $\vek{x}$ rotates by $\phi_x$. Consequently, $\langle \gamma\gamma\rangle (\vek{x})$ has a spin of 4. Being the product of $\langle \gamma\gamma\rangle (\vek{x})$ and a phase factor of $\expo{-4\ic\phi_x}$, the quantity $\xi_-(x)$ no longer depends on the polar angle of $\vek{x}$.

The relation (\ref{eq:xi+xi-}) has already taken both the statistical homogeneity and isotropy of the shear field into account and is therefore a one-dimensional relation of quantities on the real domain. The derivation of the $\xi_+-\xi_-$ relation originates from the relation between $\xi_+$ and $\xi_-$ and the convergence power spectrum $P_{\kappa}$,
\eq{
\label{eq:J0J4}
P_{\kappa}(\ell) = 2\pi\int_0^{\infty}\;\dd x\;x\;\xi_+(x)\;J_0(\ell x) = 2\pi\int_0^{\infty}\;\dd x\;x\;\xi_-(x)\;J_4(\ell x)\,.
}
Inverting one of the relations in (\ref{eq:J0J4}) one can write $\xi_+$ and $\xi_-$ in terms of each other, e.g.
\eq{
\label{eq:J0J4inv}
\xi_-(x) = \int_0^{\infty}\;\frac{\dd \ell\;\ell}{2\pi}\;J_4(\ell x)\;P_{\kappa}(\ell) = \int_0^{\infty}\;\dd y\;y\;\xi_+(y)\; \int_0^{\infty}\;{\dd \ell\;\ell}\;J_4(\ell x)\;J_0(\ell y)\,,
}
and the final form of the relation (\ref{eq:xi+xi-}) can be reached by performing the 1D Bessel integral whose result can be obtained from \citet{grad00}.

The same procedure, however, fails to work for 3-pt statistics since the corresponding Bessel integral actually consists of three integrals, and they have highly complicated dependencies on the arguments \citep[see][]{s05}. A brute force numerical evaluation of these integrals is also extremely challenging due to the oscillatory behaviour of the Bessel functions.

Since the advantage of transforming to the Fourier plane and back no longer holds for 3-pt statistics, we attempt to stay in configuration space, which at least avoids the problem of oscillatory integrals. One can see from (\ref{eq:xi+xi-}) that the result of the Bessel integral in (\ref{eq:J0J4inv}) is actually not oscillatory, as expected. 

The configuration space 3-pt shear correlator can be written as $\ba{\gamma(\vek{X_1})\gamma(\vek{X_2})\gamma(\vek{X_3})}$, with $\vek{X_i}$ being the positions on the two-dimensional (2D) plane where the shear signals are evaluated. Following the assumed statistical homogeneity of the shear field, the correlator depends only on the separations of these three positions. We choose $\vek{x_1}\equiv \vek{X_1}- \vek{X_3}$ and $\vek{x_2} \equiv \vek{X_2}- \vek{X_3}$ to be its arguments (see the leftmost sketch of Fig.\;\ref{fig:permuflip}) and write the correlator as $\langle \gamma\gamma\gamma \rangle(\vek{x_1},\vek{x_2})$. After the same procedure is applied to the 3-pt convergence correlator, the relation we are interested in will be shown to be of the form
\eq{
\label{eq:defineG}
\langle \gamma\gamma\gamma \rangle(\vek{x_1},\vek{x_2}) =  -\frac{1}{\pi^3}\;\int \dd^2 y_1 \int \dd^2 y_2\; \langle \kappa\kappa\kappa \rangle (\vek{y_1},\vek{y_2}) \;G_0(\vek{x_1}-\vek{y_1},\vek{x_2}-\vek{y_2})\,,
}
where we have defined the convolution kernel $G_0$ for which we need to find an explicit expression.

Writing the relation in the form of a convolution is motivated by the Kaiser-Squires (K-S) relation between the convergence and the shear \citep{kaiser93},
\eq{
\label{eq:KS}
\gamma(\vek{x}) = \frac{1}{\pi} \int \dd^2 y\; \kappa(\vek{y}) \; \mathcal{D}(\vek{x}-\vek{y})\,,\ \textrm{with the K-S kernel}\   \mathcal{D}(\vek{z}) = -\frac{1}{\vek{z}^{*2}}\;,
}
which yields the result (\ref{eq:defineG}) and also allows us to express the kernel $G_0$ as
 \eq{
\label{eq:gkerneldef}
G_0(\vek{a},\vek{b}) = -\int \dd^2 v\; {\mathcal{D}}(\vek{v})\;{\mathcal{D}}(\vek{v}-\vek{a})\;{\mathcal{D}}(\vek{v}-\vek{b}) = \int \dd^2 v\; \frac{1}{\vek{v}^{*2}}\;\frac{1}{(\vek{v}^*-\vek{a}^*)^2}\;\frac{1}{(\vek{v}^*-\vek{b}^*)^2}\;.
}
Here, for simplicity, we have adopted the complex notation for the K-S kernel, i.e. we have identified the 2D separation vectors with complex numbers. Throughout the text we will use the vector and complex notations interchangeably, and use $\vek{x}$ to indicate a complex quantity, $x$ for its absolute value, and $\vek{x}^*$ for its complex conjugate. 

The integral in (\ref{eq:gkerneldef}) is difficult to perform directly, so we first take a look at the more studied 2-pt case. The relation between 2-pt $\gamma$ and $\kappa$ correlation functions can be written in the same way as
\eq{
\label{eq:defineF}
\langle \gamma\gamma\rangle (\vek{x}) =  \frac{1}{\pi^2}\int \dd^2 y \; \langle \kappa\kappa \rangle (\vek{y}) \; F(\vek{x}-\vek{y}) \,,
}
with
\eq{
\label{eq:fkerneldef}
F(\vek{z}) = \int \dd^2 v\; {\mathcal{D}}(\vek{v})\;{\mathcal{D}}(\vek{v}-\vek{z}) = \int \dd^2 v\; \frac{1}{\vek{v}^{*2}}\;\frac{1}{(\vek{v}^*-\vek{z}^*)^2}\,.
}
Unlike the case of the $\xi_+-\xi_-$ relation, we have not assumed a statistically isotropic field for (\ref{eq:defineG}) or (\ref{eq:defineF}). The $\xi_+-\xi_-$ relation is actually what one should obtain after adding the assumption of isotropy to (\ref{eq:defineF}).

\subsection{The form of the convolution kernels}
\label{sec:formsofk}
Now we aim for obtaining the forms of the $F$ and $G_0$ kernels, which can be seen as the 2- and 3-pt equivalence of the K-S kernel (\ref{eq:KS}). Introducing the symbols $\partial \equiv \partial_1 + \ic \partial_2$ and $\nabla^2 \equiv \partial_1^2 + \partial_2^2 = \partial \partial^*$, the definitions of $\kappa$ and $\gamma$ read
\eq{
\label{eq:kgd2}
\kappa = \frac{1}{2} \nabla^2 \psi \,,\ \ \ \vek{\gamma} = \frac{1}{2} \partial^2 \psi \,,
}
i.e. both the convergence $\kappa$ and the shear $\vek{\gamma}$ are second-order derivatives of the deflection potential $\psi$. It is then convenient to use $\psi$ as a link between $\kappa$ and $\gamma$. Using the identities $\nabla \ln|\vek{x}| = \vek{x}/|\vek{x}|^2$ and $\nabla^2\ln|\vek{x}| = 2\pi \delta^{(2)}_{\rm D}(\vek{x})$ which hold for a 2D $\vek{x}$, one can easily verify the consistency of (\ref{eq:kgd2}) with the relation between $\psi$ and $\kappa$ \citep[e.g.][]{bartelmann01},
\eq{
\label{eq:kpsi}
\psi(\vek{x}) = \frac{1}{\pi} \int \dd^2 y \;\kappa(\vek{y}) \ln|\vek{x}-\vek{y}| \,.
}
Applying the operator $\partial^2$ on both sides of (\ref{eq:kpsi}) and taking (\ref{eq:kgd2}) into account, one reaches the K-S relation (\ref{eq:KS}), since $\cal{D}(\vek{z}) =$$\partial^2\ln|\vek{z}|$.

The same procedure can be generalized to second-order statistics. The 2-pt equivalence of (\ref{eq:kpsi}) is
\eq{
\label{eq:2kpsi}
\ba{\psi(\vek{x_1}) \psi(\vek{x_2})} = \frac{1}{\pi^2} \int \dd^2 y_1 \; \ln|\vek{x_1}-\vek{y_1}| \; \int \dd^2 y_2 \; \ln|\vek{x_2}-\vek{y_2}|\;  \ba{ \kappa(\vek{y_1}) \kappa(\vek{y_2})}\,. 
}
Using the statistical homogeneity of the $\kappa$ field, and re-defining the integration variables, (\ref{eq:2kpsi}) reduces to
\eqs{
\label{eq:2kpsi2}
\ba{\psi(\vek{x_1}) \psi(\vek{x_2})}& =  \frac{1}{\pi^2} \int \dd^2 y \;\ba{ \kappa\kappa}(\vek{y}) \;  \int \dd^2 u \;  \ln|\vek{u}|  \ln|\vek{x_1} - \vek{x_2} - \vek{y} - \vek{u}|\\
& =  \frac{1}{\pi^2} \int \dd^2 y \;\ba{ \kappa\kappa}(\vek{y}) \; \mathcal{F'}\br{ \vek{x_1} - \vek{x_2} - \vek{y}}\,,
}
where we have defined 
\eq{
 \mathcal{F'}\br{\vek{z}} =  \int \dd^2 u \;  \ln \bs{\vek{u}}  \ln\bs{\vek{z}-\vek{u}} \,.
}
Obviously, $\mathcal{F'}$ is infinite at every $\vek{z}$, which is related to the fact that $\psi$ is defined only up to an additive constant. However, we shall only need the derivatives of $\mathcal{F'}$. So we define
\eq{
\label{eq:xx1}
 \mathcal{F}\br{\vek{z}}  = \mathcal{F'}\br{\vek{z}}- \mathcal{F'}\br{\vek{0}} = \int \dd^2 u \;  \ln \bs{\vek{u}}  \ln\br{\frac{\bs{\vek{z}-\vek{u}}}{\bs{\vek{u}} }} \,,
}
and will use $\mathcal{F}$ and $\mathcal{F'}$ interchangeably.
Let $\varphi$ denote the angle between $\vek{u}$ and $\vek{z}$, (\ref{eq:xx1}) can be rewritten as 
\eq{
\mathcal{F}\br{\vek{z}} = \frac{1}{2} \int_0^{\infty} \dd u \; u\;  \ln u \; \int_0^{2\pi} \dd \varphi\; \ln\br{1-\frac{2|\vek{z}|}{u} \cos \varphi + \frac{|\vek{z}|^2}{u^2}}\,. 
}
The integral over $\varphi$ yields zero if $|\vek{z}|<u$, and $4\pi \ln(|\vek{z}|/u)$ otherwise. Thus
\eq{
\label{eq:16}
\mathcal{F}\br{\vek{z}} = 2\pi \int_0^{|\vek{z}|} \dd u \; u\;  \ln u \;\ln(|\vek{z}|/u) = \frac{\pi}{2} |\vek{z}|^2 \br{\ln|\vek{z}|-1}\,.
}

We are now ready to apply differential operators to (\ref{eq:2kpsi2}) to get the relations of 2-pt shear and convergence statistics. As a consistency check, we first apply two $\nabla^2$ operators to (\ref{eq:2kpsi2}), one acting on $\vek{x_1}$ and the other on $\vek{x_2}$. According to (\ref{eq:kgd2}), this turns the l.h.s. of (\ref{eq:2kpsi2}) into $4 \ba{\kappa(\vek{x_1})\kappa(\vek{x_2})}$. On the r.h.s. of (\ref{eq:2kpsi2}) the operators act exclusively on $\mathcal{F}$, 
\eq{
\label{eq:nabnabF}
\nabla^2_{{x_1}}\nabla^2_{{x_2}}\mathcal{F}\br{ \vek{x_1} - \vek{x_2} - \vek{y}} = \nabla^2\nabla^2  \mathcal{F}\br{\vek{z}}= \nabla^2\br{2\pi\ln|\vek{z}|} = 4\pi^2 \delta^{(2)}_{\rm D}(\vek{z})\,,
}
with $\vek{z}=\vek{x_1} - \vek{x_2} - \vek{y}$ here. Using (\ref{eq:nabnabF}), one easily sees that the r.h.s. of (\ref{eq:2kpsi2}) after the operation gives $4 \ba{ \kappa\kappa}(\vek{x_1}-\vek{x_2})$, which is equivalent to $4 \ba{\kappa(\vek{x_1})\kappa(\vek{x_2})}$ under the assumption of statistical homogeneity of the $\kappa$ field.  

Now we apply the operator $\partial^2_{{x_1}}\partial^2_{{x_2}}/4$ on (\ref{eq:2kpsi2}), which turns the l.h.s. of (\ref{eq:2kpsi2}) into $\ba{\gamma(\vek{x_1})\gamma(\vek{x_2})}$. On the r.h.s. the operation again acts only on $\mathcal{F}$, 
\eq{
\frac{1}{4}\partial^2_{{x_1}}\partial^2_{{x_2}} \mathcal{F}\br{ \vek{x_1} - \vek{x_2} - \vek{y}} =\frac{1}{4} \partial^4  \mathcal{F}\br{\vek{z}}\,,
}
also with $\vek{z}=\vek{x_1} - \vek{x_2} - \vek{y}$.
Remembering the definition of the kernel $F$ (\ref{eq:defineF}), this leads to
\eq{
\label{eq:Fform}
F(\vek{z}) = \int \dd^2 v\; \frac{1}{\vek{v}^{*2}}\;\frac{1}{(\vek{v}^*-\vek{z}^*)^2} =\frac{1}{4} \partial^4  \mathcal{F}\br{\vek{z}} = 2\pi \frac{\vek{z}}{\vek{z}^{*3}}\,.
}

For the 3-pt kernel $G_0$ we split the integral in (\ref{eq:gkerneldef}) into
\eqs{
\label{eq:spplit}
 &\int \dd^2 v\; \frac{1}{\vek{v}^{*2}}\;\frac{1}{(\vek{v}^*-\vek{a}^*)^2}\;\frac{1}{(\vek{v}^*-\vek{b}^*)^2} \\
= \;& \frac{1}{(\vek{a}^*-\vek{b}^*)^2} \int \dd^2 v\; \frac{1}{\vek{v}^{*2}}\;\bb{ \frac{1}{(\vek{v}^*-\vek{a}^*)^2} + \frac{1}{(\vek{v}^*-\vek{b}^*)^2}} - \frac{2}{(\vek{a}^*-\vek{b}^*)^3} \int \dd^2 v\; \frac{1}{\vek{v}^{*2}}\;\bb{ \frac{1}{\vek{v}^*-\vek{a}^*} - \frac{1}{\vek{v}^*-\vek{b}^*}}\,,
}
where we have assumed $\vek{a}\neq\vek{b}$.
From (\ref{eq:Fform}) as well as
\eq{
\int \dd^2 v\; \frac{1}{\vek{v}^{*2}}\;\frac{1}{\vek{v}^*-\vek{z}^*} =\frac{1}{2} \partial^{3} \mathcal{F}\br{\vek{z}}  =  -\pi \frac{\vek{z}}{\vek{z}^{*2}}\;,
}
we obtain
\eq{
\label{eq:G2p}
G_0(\vek{a},\vek{b}) = \int \dd^2 v\; \frac{1}{\vek{v}^{*2}}\;\frac{1}{(\vek{v}^*-\vek{a}^*)^2}\;\frac{1}{(\vek{v}^*-\vek{b}^*)^2} =  \frac{2\pi}{(\vek{a}^*-\vek{b}^*)^2} \br{\frac{\vek{a}}{\vek{a}^{*3}}+\frac{\vek{b}}{\vek{b}^{*3}}} + \frac{2\pi}{(\vek{a}^*-\vek{b}^*)^3}\br{\frac{\vek{a}}{\vek{a}^{*2}}-\frac{\vek{b}}{\vek{b}^{*2}}}  \;.
}

The forms of the kernels (\ref{eq:Fform}) and (\ref{eq:G2p}) hold rigorously outside their singularities (at $\vek{z}=0$ for $F$; at $\vek{a}=0$, $\vek{b}=0$, and $\vek{a}=\vek{b}$ for $G_0$). One may wonder if additional delta functions exist at these singularities. We will show in Sect.\;\ref{sec:consistency} that this is not the case.

The method we used to derive the forms of the kernels (\ref{eq:Fform}) and (\ref{eq:G2p}) also allows one to derive the relations between other correlation functions of weak lensing quantities in a systematic way. We present explicit forms of some of the relations in Appendix\;\ref{sec:mixcor}.

\subsection{The relations}

To summarize, we have obtained:
\eq{
\label{eq:Frelation}
\langle \gamma\gamma\rangle (\vek{x}) =  \frac{2}{\pi}\int \dd^2 y \; \langle \kappa\kappa \rangle (\vek{y}) \; \frac{\vek{y}-\vek{x}}{(\vek{y}^*-\vek{x}^*)^3} \,,
}
and 
\eqs{
\label{eq:Grelation}
\langle \gamma\gamma\gamma \rangle(\vek{x_1},\vek{x_2})& =  -\frac{2}{\pi^2}\;\int \dd^2 y_1 \int \dd^2 y_2\; \langle \kappa\kappa\kappa \rangle (\vek{y_1},\vek{y_2}) \; \Bigg[ \frac{1}{{(\vek{y_1}^*-\vek{x_1}^*-\vek{y_2}^*+\vek{x_2}^*)^2}} \br{\frac{\vek{y_1}-\vek{x_1}}{(\vek{y_1}^*-\vek{x_1}^*)^{3}}+\frac{\vek{y_2}-\vek{x_2}}{(\vek{y_2}^*-\vek{x_2}^*)^{3}}}\\
& + \frac{1}{(\vek{y_1}^*-\vek{x_1}^*-\vek{y_2}^*+\vek{x_2}^*)^3}\br{\frac{\vek{y_1}-\vek{x_1}}{(\vek{y_1}^*-\vek{x_1}^*)^{2}}-\frac{\vek{y_2}-\vek{x_2}}{(\vek{y_2}^*-\vek{x_2}^*)^{2}}} \Bigg]\,.
}

In these relations we have applied the statistical homogeneity of the convergence field, but not the statistical isotropy. Making use of the latter, one can derive the $\xi_+-\xi_-$ relation from (\ref{eq:Frelation}), as will be shown in Sect.\;\ref{sec:consisxixi}.

\section{Consistency checks}
\label{sec:consistency}
\subsection{The case of uniform $\kappa$}

There is a physical condition which will directly serve as a test of the $\kappa\; -\; \gamma$ relations (\ref{eq:KS}), (\ref{eq:Frelation}) and (\ref{eq:Grelation}).
At the 1-pt level, for the K-S relation, a uniform convergence field does not result in any shear.
At the 2- and 3-pt level, the physical condition could be that a uniform $\langle \kappa\kappa \rangle$ ($\langle \kappa\kappa\kappa \rangle$) field leads to a vanishing shear correlation $\langle \gamma\gamma \rangle$ ($\langle \gamma\gamma\gamma \rangle$). 

One can easily see that both the $F$ and $G_0$ kernel we obtained satisfy this condition. If there are additional terms at the singularities of the kernels which contribute to the integral, a non-zero $\langle \gamma\gamma \rangle$ ($\langle \gamma\gamma\gamma \rangle$) term would be generated and the condition would not be satisfied anymore. Thus we argue that the expressions (\ref{eq:Frelation}) and (\ref{eq:Grelation}) are already complete.

\subsection{Consistency with the $\xi_+ - \xi_-$ relation}
\label{sec:consisxixi}
Now we consider whether (\ref{eq:Frelation}) is consistent with the $\xi_+ - \xi_-$ relation (\ref{eq:xi+xi-}), which can be regarded as the isotropic form of (\ref{eq:Frelation}). That the two relations are consistent is equivalent to 
\eq{
\label{eq:2dphiint}
\int_0^{2\pi} \dd\phi_y  \; \frac{\vek{y}-\vek{x}}{(\vek{y}^*-\vek{x}^*)^3} =  \frac{\pi}{2} \;\expo{4\ic \phi_x} \;  \bb{\frac{4 x^2-12 y^2}{x^4} H(x-y) + \frac{\delta(x-y)}{x} } \,.
}

To verify that (\ref{eq:2dphiint}) indeed holds, we attempt to solve the $\phi_y$-integral on the l.h.s., 
\eq{
\label{eq:phiyint}
 \int_0^{2\pi} \dd\phi_y  \; \frac{\vek{y}-\vek{x}}{(\vek{y}^*-\vek{x}^*)^3} = \expo{4\ic \phi_x} \int_0^{2\pi} \dd\phi \; \frac{y\expo{\ic\phi}-x}{(y\expo{-\ic\phi}-x)^3}\,,
}
where $\phi = \phi_y- \phi_x$ has been defined. The $\phi$-integral can be carried out using the residual theorem, yielding $2\pi (x^2- 3 y^2)/x^4$ when $x>y$ and zero when $x<y$. One can see that this result corresponds to the Heaviside function on the r.h.s. of (\ref{eq:2dphiint}).  

At the singularity $\vek{x}=\vek{y}$ the $\phi$-integral is not well defined, which means one cannot rule out the existence of additional delta function at $x=y$ in the result of the $\phi$-integral. This ambiguity can again be eliminated by using the physical condition \textit{`a uniform $\langle \kappa\kappa \rangle$ field leads to a null shear correlation $\langle \gamma\gamma \rangle$'}, which translates to \textit{`a constant $\xi_+$ yields vanishing $\xi_-$'} here. In this case, a delta function is indeed required to satisfy this condition, and the prefactor of the delta function can be determined to be $\pi/2x$, in consistency with (\ref{eq:2dphiint}).

\subsubsection{K-S relation and its isotropic form}
A similar consistency exists between the K-S relation and its isotropic form. As both forms are already well-known, they can serve as a further support for our argument.

For an axisymmetric distribution of matter, i.e. $\kappa(\vek{x}) = \kappa(x)$, the following relation is established between the shear and the convergence \citep[see e.g.][]{s92}
\eq{
\gamma(\vek{x}) = \bb{\kappa(x) - \bar{\kappa}(x)} \; \expo{2 \ic\phi_x}\,,
}
with $\bar{\kappa}$ defined as
\eq{
\bar{\kappa}(x) := \frac{2}{x^2} \int_0^x y\;\dd y\; \kappa(y)\,.
}
This is equivalent to 
\eq{
\label{eq:ksaxiasym}
\gamma(\vek{x}) = -\frac{1}{\vek{x}^{*2}} \int y\; \dd y \;\kappa(y) \;\bb{2H(x-y) - x \;\delta^{(1)}_{\rm D}(x-y)}\,.
}
In the case of a uniform convergence field $\kappa(x)=$ const., one can see that the integral of the Heaviside function and the delta function parts cancel each other. 

The similarity between (\ref{eq:ksaxiasym}) and the $\xi_+ - \xi_-$ relation (\ref{eq:xi+xi-}) is remarkable: they both have integrals of a Heaviside function part and a delta function part which cancel each other for constant $\kappa$ and $\ba{\kappa\kappa}$, respectively, and the 2D correspondences of both do not have an additional delta function at their singularities.

\subsection{Fourier transformations}
\label{sec:fouriertransfs}
In the Fourier plane the relation between the shear and the convergence has a simple form. With $\tilde{\vek{g}}$ denoting the Fourier transformation of the quantity $\vek{g}$, it reads
\eq{
\label{eq:Fkgd2}
\tilde{\gamma}(\vek{\ell}) = \expo{2\ic\beta}\;\tilde{\kappa}(\vek{\ell})\,,\ \ \textrm{for}\ \vek{\ell}\ne \vek{0}\,,
}
with $\beta$ being the polar angle of the wave vector $\vek{\ell}$. 
The identity (\ref{eq:Fkgd2}) can be obtained by relating the Fourier transforms of the definitions (\ref{eq:kgd2}) of $\gamma$ and $\kappa$. It directly reflects the fact that $\gamma$ is spin-2 while $\kappa$ is spin-0, and leads to the well-known result $P_{\gamma} = P_{\kappa}$. Comparing (\ref{eq:Fkgd2}) to the configuration space relation (\ref{eq:KS}) which means $\gamma = \kappa * \cal{D}/\pi$, using the convolution theorem, one can see that $\expo{2\ic\beta} = \tilde{\cal{D}}/\pi$ for $\vek{\ell}\ne \vek{0}$. 

The Fourier plane correspondences of (\ref{eq:Frelation}) and (\ref{eq:Grelation}) are also readily obtainable from the identity (\ref{eq:Fkgd2}), as
\eq{
\label{eq:FourierF}
\langle \tilde{\gamma}\tilde{\gamma} \rangle(\vek{\ell})=\expo{4\ic \beta}\; \langle \tilde{\kappa}\tilde{\kappa} \rangle(\vek{\ell})\,,\ \ \textrm{for}\ \vek{\ell}\ne \vek{0}\,,
}
and
\eq{
\label{eq:FourierG}
\langle \tilde{\gamma}\tilde{\gamma}\tilde{\gamma} \rangle(\vek{\ell_1},\vek{\ell_2},\vek{\ell_3})=\expo{2\ic\br{\beta_1+\beta_2+\beta_3}}\; \langle \tilde{\kappa}\tilde{\kappa}\tilde{\kappa} \rangle(\vek{\ell_1},\vek{\ell_2},\vek{\ell_3})\,,\ \ \textrm{for}\ \vek{\ell_1}, \vek{\ell_2}, \vek{\ell_3} \ne \vek{0}\,,
}
with $\beta_i$ denoting the polar angle of $\vek{\ell_i}$. 
These equations show that $\expo{4\ic \beta} = \tilde{F}(\vek{\ell}) /\pi^2$ for $\vek{\ell} \ne \vek{0}$, and that $\expo{2\ic\br{\beta_1+\beta_2+\beta_3}} = -\tilde{G_0}(\vek{\ell_1},\vek{\ell_2}) /\pi^3$ for $\vek{\ell_1} \ne \vek{0} \ne \vek{\ell_2}$ and $\vek{\ell_3} = -\vek{\ell_1}-\vek{\ell_2} \ne 0$, since $F /\pi^2$ and $-G_0 /\pi^3$ are the convolution kernels for the configuration space relations by their definitions. 

In Appendix\,\ref{app:1} we show explicitly that the Fourier transforms of $F/\pi^2$ and $-G_0/\pi^3$, with $F$ and $G_0$ given in (\ref{eq:Fform}) and (\ref{eq:G2p}), are indeed $\expo{4\ic \beta}$ and $\expo{2\ic\br{\beta_1+\beta_2+\beta_3}}$, respectively. However one cannot obtain the forms of $F$ and $G_0$ kernels simply through inverse Fourier transforming the phase factors $\expo{4\ic\beta}$ and $\expo{2\ic\br{\beta_1+\beta_2+\beta_3}}$. This is due to the fact that the Fourier inversion theorem is valid strictly only for square-integrable functions, which is not the case for the phase factors. The same situation occurs for the K-S kernel $\cal{D}$.

\section{The other $\gamma$3PCFs}
Until now we have considered only the 3PCF of shear itself $\ba{\gamma(\vek{X_1})\gamma(\vek{X_2})\gamma(\vek{X_3})}$, which is one of the four independent possible combinations considering that $\gamma$ is a complex quantity. The other three are $\ba{\gamma^*(\vek{X_1})\gamma(\vek{X_2})\gamma(\vek{X_3})}$, $\ba{\gamma(\vek{X_1})\gamma^*(\vek{X_2})\gamma(\vek{X_3})}$, and $\ba{\gamma(\vek{X_1})\gamma(\vek{X_2})\gamma^*(\vek{X_3})}$, according to the choice made in \citet{s03a}. Following \citet{s05}, we denote these four $\gamma$3PCFs by $\Gamma^{(i)}_{\rm cart}(\vek{x_1},\vek{x_2})$ ($i=0,1,2,3$), with `cart' emphasizing that the shear is measured in Cartesian coordinates, $\Gamma^{(0)}_{\rm cart} \equiv \ba{\gamma\gamma\gamma}$, and $\Gamma^{(i)}_{\rm cart}$ ($i=1,2,3$) corresponding to the $\gamma$3PCF with $\gamma^*$ at position $\vek{X_i}$. Since we have considered statistical homogeneity of the shear field, the $\Gamma_{\rm cart}$'s depend only on the separation vectors of the position \vek{X_1}, \vek{X_2}, and \vek{X_3}. The other $\gamma$3PCFs, i.e. those with two or three $\gamma^*$'s, can be obtained by taking the complex conjugate of the $\Gamma_{\rm cart}$'s. 

\begin{figure}[h]
\centering
\includegraphics[width=17cm]{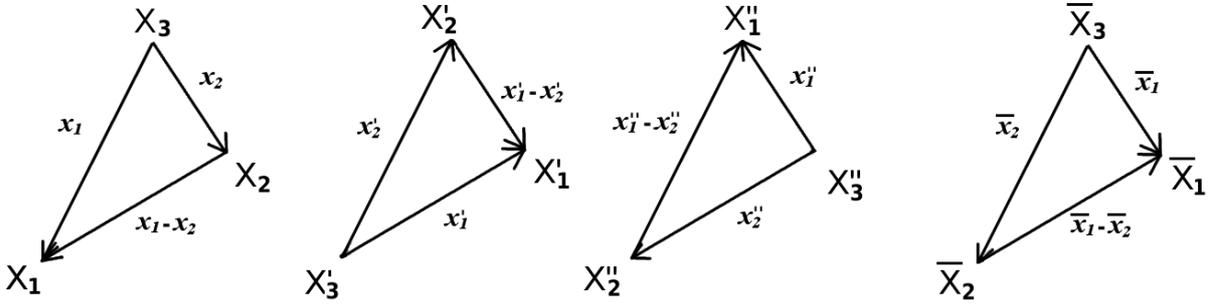}
\centering
\caption{Definition of the geometry of a triangle (the leftmost sketch) and how it changes under permutations (the first three sketches from the left) and flip (the leftmost and the rightmost sketch) of the vertices.}
\label{fig:permuflip}
\end{figure}

Note that $\Gamma^{(1)}_{\rm cart}(\vek{x_1},\vek{x_2}) \equiv \langle \gamma^*\gamma\gamma \rangle(\vek{x_1},\vek{x_2})$, $\Gamma^{(2)}_{\rm cart}(\vek{x_1},\vek{x_2}) \equiv \langle \gamma\gamma^*\gamma \rangle(\vek{x_1},\vek{x_2})$, and $\Gamma^{(3)}_{\rm cart}(\vek{x_1},\vek{x_2}) \equiv \langle \gamma\gamma\gamma^* \rangle(\vek{x_1},\vek{x_2})$ are different functions, since $\vek{x_1}$ (\vek{x_2}) is defined to be the difference of the positions of the first (second) and the third $\gamma$ in the bracket. Due to the same reason, they can be transformed into each other through permutations and flips of the vertices of the triangle formed by their arguments (see Fig.\;\ref{fig:permuflip}), and thus are not independent if argument permutations and flips are allowed. As an example, one has
\eqs{
\label{eq:symmetry}
 &\ba{\gamma^*(\vek{X_1})\gamma(\vek{X_2})\gamma(\vek{X_3})} \equiv \Gamma^{(1)}_{\rm cart}(\vek{x_1},\vek{x_2})\\
=&\ba{\gamma^*(\vek{X_1})\gamma(\vek{X_3})\gamma(\vek{X_2})} \equiv \Gamma^{(1)}_{\rm cart}(\vek{x_1}-\vek{x_2},-\vek{x_2})\\
=&\ba{\gamma(\vek{X_2})\gamma^*(\vek{X_1})\gamma(\vek{X_3})} \equiv \Gamma^{(2)}_{\rm cart}(\vek{x_2},\vek{x_1})\\
=&\ba{\gamma(\vek{X_3})\gamma^*(\vek{X_1})\gamma(\vek{X_2})} \equiv \Gamma^{(2)}_{\rm cart}(-\vek{x_2},\vek{x_1}-\vek{x_2})\\
=&\ba{\gamma(\vek{X_2})\gamma(\vek{X_3})\gamma^*(\vek{X_1})} \equiv \Gamma^{(3)}_{\rm cart}(\vek{x_2}-\vek{x_1},-\vek{x_1})\\
=&\ba{\gamma(\vek{X_3})\gamma(\vek{X_2})\gamma^*(\vek{X_1})} \equiv \Gamma^{(3)}_{\rm cart}(-\vek{x_1},\vek{x_2}-\vek{x_1})\,,
}
where different lines correspond to different ways of labeling the same triangle with side lengths $x_1$, $x_2$, and $|\vek{x_1}-\vek{x_2}|$. The same permutations and flips also reveal the inherent symmetry of $\Gamma^{(0)}_{\rm cart}$, 
\eq{
\label{eq:sym0}
\Gamma^{(0)}_{\rm cart}(\vek{x_1},\vek{x_2}) = \Gamma^{(0)}_{\rm cart}(\vek{x_1}-\vek{x_2},-\vek{x_2}) = \Gamma^{(0)}_{\rm cart}(\vek{x_2},\vek{x_1}) = \Gamma^{(0)}_{\rm cart}(-\vek{x_2},\vek{x_1}-\vek{x_2}) = \Gamma^{(0)}_{\rm cart}(\vek{x_2}-\vek{x_1},-\vek{x_1}) = \Gamma^{(0)}_{\rm cart}(-\vek{x_1},\vek{x_2}-\vek{x_1})\,.
}

In the case that the shear is measured relative to a center of the triangle, $\Gamma^{(i)}_{\rm cart}$ transforms to $\Gamma^{(i)}$, the natural components of the $\gamma$3PCF as defined in \citet{s03a}. For a general triangle configuration, all four $\Gamma_{\rm cart}$'s are expected to be non-zero, thus all of them should be used to exploit the full 3-pt information of cosmic shear. 

Before relating the other $\Gamma_{\rm cart}$'s to the $\kappa$3PCFs, we extend $\kappa$ to a complex quantity $\kappa = \kappa^{\rm{E}}+\ic \kappa^{\rm{B}}$. Although the physical convergence is a real quantity, the convergence field corresponding to the measured shear signals can have an imaginary part due to e.g. systematical errors and noise. The shear component which corresponds to this unphysical imaginary part of the convergence field is identified as the B-mode, on which we will elaborate more in Sect.\;\ref{sec:EB}. When taking the B-mode into consideration, the 3-pt correlation functions of the convergence field can be written as
\eqs{
\label{eq:KEB}
K^{(0)} \equiv \langle \kappa\kappa\kappa \rangle &= \langle \kappa^{\rm{E}}\kappa^{\rm{E}}\kappa^{\rm{E}} \rangle +  \ic \langle \kappa^{\rm{B}}\kappa^{\rm{E}}\kappa^{\rm{E}} \rangle+  \ic \langle \kappa^{\rm{E}}\kappa^{\rm{B}}\kappa^{\rm{E}} \rangle+  \ic \langle \kappa^{\rm{E}}\kappa^{\rm{E}}\kappa^{\rm{B}} \rangle -  \langle \kappa^{\rm{E}}\kappa^{\rm{B}}\kappa^{\rm{B}} \rangle-  \langle \kappa^{\rm{B}}\kappa^{\rm{E}}\kappa^{\rm{B}} \rangle-  \langle \kappa^{\rm{B}}\kappa^{\rm{B}}\kappa^{\rm{E}} \rangle - \ic  \langle \kappa^{\rm{B}}\kappa^{\rm{B}}\kappa^{\rm{B}} \rangle \,,\\
K^{(1)} \equiv \langle \kappa^*\kappa\kappa \rangle &= \langle \kappa^{\rm{E}}\kappa^{\rm{E}}\kappa^{\rm{E}} \rangle - \ic \langle \kappa^{\rm{B}}\kappa^{\rm{E}}\kappa^{\rm{E}} \rangle+ \ic \langle \kappa^{\rm{E}}\kappa^{\rm{B}}\kappa^{\rm{E}} \rangle + \ic \langle \kappa^{\rm{E}}\kappa^{\rm{E}}\kappa^{\rm{B}} \rangle - \langle \kappa^{\rm{E}}\kappa^{\rm{B}}\kappa^{\rm{B}} \rangle + \langle \kappa^{\rm{B}}\kappa^{\rm{E}}\kappa^{\rm{B}} \rangle + \langle \kappa^{\rm{B}}\kappa^{\rm{B}}\kappa^{\rm{E}} \rangle + \ic  \langle \kappa^{\rm{B}}\kappa^{\rm{B}}\kappa^{\rm{B}} \rangle \,,\\
K^{(2)} \equiv \langle \kappa\kappa^*\kappa \rangle &= \langle \kappa^{\rm{E}}\kappa^{\rm{E}}\kappa^{\rm{E}} \rangle + \ic \langle \kappa^{\rm{B}}\kappa^{\rm{E}}\kappa^{\rm{E}} \rangle- \ic \langle \kappa^{\rm{E}}\kappa^{\rm{B}}\kappa^{\rm{E}} \rangle + \ic \langle \kappa^{\rm{E}}\kappa^{\rm{E}}\kappa^{\rm{B}} \rangle + \langle \kappa^{\rm{E}}\kappa^{\rm{B}}\kappa^{\rm{B}} \rangle - \langle \kappa^{\rm{B}}\kappa^{\rm{E}}\kappa^{\rm{B}} \rangle + \langle \kappa^{\rm{B}}\kappa^{\rm{B}}\kappa^{\rm{E}} \rangle + \ic  \langle \kappa^{\rm{B}}\kappa^{\rm{B}}\kappa^{\rm{B}} \rangle \,,\\
K^{(3)} \equiv \langle \kappa\kappa\kappa^* \rangle &= \langle \kappa^{\rm{E}}\kappa^{\rm{E}}\kappa^{\rm{E}} \rangle + \ic \langle \kappa^{\rm{B}}\kappa^{\rm{E}}\kappa^{\rm{E}} \rangle+ \ic \langle \kappa^{\rm{E}}\kappa^{\rm{B}}\kappa^{\rm{E}} \rangle - \ic \langle \kappa^{\rm{E}}\kappa^{\rm{E}}\kappa^{\rm{B}} \rangle + \langle \kappa^{\rm{E}}\kappa^{\rm{B}}\kappa^{\rm{B}} \rangle + \langle \kappa^{\rm{B}}\kappa^{\rm{E}}\kappa^{\rm{B}} \rangle - \langle \kappa^{\rm{B}}\kappa^{\rm{B}}\kappa^{\rm{E}} \rangle + \ic  \langle \kappa^{\rm{B}}\kappa^{\rm{B}}\kappa^{\rm{B}} \rangle \,.
}
Apart from the E-mode $\langle \kappa^{\rm{E}}\kappa^{\rm{E}}\kappa^{\rm{E}} \rangle$ term, there are still additional B-mode contributions to the real parts of the $K$'s, namely $\langle \kappa^{\rm{E}}\kappa^{\rm{B}}\kappa^{\rm{B}} \rangle$, $\langle \kappa^{\rm{B}}\kappa^{\rm{E}}\kappa^{\rm{B}} \rangle$, and $\langle \kappa^{\rm{B}}\kappa^{\rm{B}}\kappa^{\rm{E}} \rangle$. The imaginary part of the $K$'s are composed of the parity violating terms which are expected to vanish due to parity symmetry \citep{s03}. The property of $K^{(i)}$ under permutations and flips of the vertices of the triangle formed by their arguments is the same as that of $\Gamma^{(i)}_{\rm cart}$. 

Similar to (\ref{eq:defineG}), the relations between $\Gamma^{(i)}_{\rm cart}$ and $K^{(i)}$ for $i=1,2,3$ can be written as   
\eq{
\label{eq:defineG1}
\Gamma^{(1)}_{\rm cart}(\vek{x_1},\vek{x_2})  =  -\frac{1}{\pi^3}\;\int \dd^2 y_1 \int \dd^2 y_2\; K^{(1)}(\vek{y_1},\vek{y_2}) \;G_1(\vek{x_1}-\vek{y_1},\vek{x_2}-\vek{y_2})\,,
}
\eq{
\label{eq:defineG2}
\Gamma^{(2)}_{\rm cart}(\vek{x_1},\vek{x_2})  =  -\frac{1}{\pi^3}\;\int \dd^2 y_1 \int \dd^2 y_2\; K^{(2)}(\vek{y_1},\vek{y_2}) \;G_2(\vek{x_1}-\vek{y_1},\vek{x_2}-\vek{y_2})\,,
}
and
\eq{
\label{eq:defineG3}
\Gamma^{(3)}_{\rm cart}(\vek{x_1},\vek{x_2})  =  -\frac{1}{\pi^3}\;\int \dd^2 y_1 \int \dd^2 y_2\; K^{(3)}(\vek{y_1},\vek{y_2}) \;G_3(\vek{x_1}-\vek{y_1},\vek{x_2}-\vek{y_2})\,,
}
where the convolution kernels $G_1$, $G_2$, and $G_3$ have been defined. Again with the aid of the K-S relation, we can write these convolution kernels as 
\eqs{
\label{eq:G1integralf}
G_1(\vek{a},\vek{b}) = -\int \dd^2 v\; {\mathcal{D}}(\vek{v})\;{\mathcal{D}^*}(\vek{v}-\vek{a})\;{\mathcal{D}}(\vek{v}-\vek{b}) &= \int \dd^2 v\; \frac{1}{\vek{v}^{*2}}\;\frac{1}{(\vek{v}-\vek{a})^2}\;\frac{1}{(\vek{v}^*-\vek{b}^*)^2}\\
&= \int \dd^2 v\; \frac{1}{\vek{v}^{2}}\;\frac{1}{(\vek{v}^*+\vek{a}^*)^2}\;\frac{1}{(\vek{v}^*+\vek{a}^*-\vek{b}^*)^2}\;,
}
\eqs{
\label{eq:G2integralf}
G_2(\vek{a},\vek{b}) = -\int \dd^2 v\; {\mathcal{D}}(\vek{v})\;{\mathcal{D}}(\vek{v}-\vek{a})\;{\mathcal{D}^*}(\vek{v}-\vek{b}) &= \int \dd^2 v\; \frac{1}{\vek{v}^{*2}}\;\frac{1}{(\vek{v}^*-\vek{a}^*)^2}\;\frac{1}{(\vek{v}-\vek{b})^2}\\
&=\int \dd^2 v\; \frac{1}{\vek{v}^{2}}\;\frac{1}{(\vek{v}^*-\vek{a}^*+\vek{b}^*)^2}\;\frac{1}{(\vek{v}^*+\vek{b}^*)^2} \;,
}
and
\eq{
\label{eq:G3integralf}
G_3(\vek{a},\vek{b}) = -\int \dd^2 v\; {\mathcal{D}^*}(\vek{v})\;{\mathcal{D}}(\vek{v}-\vek{a})\;{\mathcal{D}}(\vek{v}-\vek{b}) = \int \dd^2 v\; \frac{1}{\vek{v}^{2}}\;\frac{1}{(\vek{v}^*-\vek{a}^*)^2}\;\frac{1}{(\vek{v}^*-\vek{b}^*)^2}\;.
}

When $\vek{a} \neq \vek{b}$, the product of the three terms in the integrand of (\ref{eq:G3integralf}) can be split into products of two, as 
\eq{
\frac{1}{\vek{v}^{2}}\;\frac{1}{(\vek{v}^*-\vek{a}^*)^2}\;\frac{1}{(\vek{v}^*-\vek{b}^*)^2} = \frac{1}{(\vek{a}^*-\vek{b}^*)^2} \frac{1}{\vek{v}^{2}}\bb{ \frac{1}{(\vek{v}^*-\vek{a}^*)^2} + \frac{1}{(\vek{v}^*-\vek{b}^*)^2}} - \frac{2}{(\vek{a}^*-\vek{b}^*)^3} \frac{1}{\vek{v}^{2}}\bb{ \frac{1}{\vek{v}^*-\vek{a}^*} - \frac{1}{\vek{v}^*-\vek{b}^*}} \,.
}
These terms are also obtainable from doing derivatives to the kernel $\mathcal{F}$, 
\eq{
\int \dd^2 v\; \frac{1}{\vek{v}^{2}} \frac{1}{(\vek{v}^*-\vek{a}^*)^2} = \frac{1}{4} \partial^2 \partial^{*2} \mathcal{F}(\vek{a}) = \pi^2 \delta^{(2)}(\vek{a})\,,
}
\eq{
\int \dd^2 v\; \frac{1}{\vek{v}^{2}} \frac{1}{\vek{v}^*-\vek{a}^*} = \frac{1}{2} \partial \partial^{*2} \mathcal{F}(\vek{a}) = \frac{\pi}{\vek{a}}\,.
}
This way we obtain the form of the convolution kernel $G_3$. The forms for the kernel $G_1$ and $G_2$ can be obtained likewise. The results are
\eq{
\label{eq:G12p}
G_1(\vek{a},\vek{b}) = \frac{\pi^2}{\vek{b}^{*2}} \bb{\delta^{(2)}_{\rm D}(\vek{a}) + \delta^{(2)}_{\rm D}(\vek{b}-\vek{a})} - \frac{2\pi}{\vek{b}^{*3}} \br{\frac{1}{\vek{a}}+\frac{1}{\vek{b}-\vek{a}}}   \;,
}
\eq{
\label{eq:G22p}
G_2(\vek{a},\vek{b}) = \frac{\pi^2}{\vek{a}^{*2}} \bb{\delta^{(2)}_{\rm D}(\vek{a}-\vek{b}) + \delta^{(2)}_{\rm D}(\vek{b})} -\frac{2\pi}{\vek{a}^{*3}} \br{\frac{1}{\vek{a}-\vek{b}}+\frac{1}{\vek{b}}}   \;,
}
\eq{
\label{eq:G32p}
G_3(\vek{a},\vek{b}) = \frac{\pi^2}{(\vek{a}^*-\vek{b}^*)^2} \bb{\delta^{(2)}_{\rm D}(\vek{a}) + \delta^{(2)}_{\rm D}(\vek{b})} -\frac{2\pi}{(\vek{a}^*-\vek{b}^*)^3} \br{\frac{1}{\vek{a}}-\frac{1}{\vek{b}}}   \;.
}

The symmetries in the $\Gamma_{\rm cart}$'s and $K$'s are also reflected in the $G$ kernels. One can verify that $G_2(\vek{a},\vek{b}) = G_1(\vek{b}-\vek{a},-\vek{a})$, $G_3(\vek{a},\vek{b}) = G_1(-\vek{b},\vek{a}-\vek{b})$ as results of the symmetry under permutations, and $G_2(\vek{a},\vek{b}) = G_1(\vek{b},\vek{a})$, $G_3(\vek{a},\vek{b}) = G_3(\vek{b},\vek{a})$ as results of the symmetry under flips, in consistency with (\ref{eq:symmetry}). Similarly, one has $G_0(\vek{a},\vek{b}) = G_0(\vek{b}-\vek{a},-\vek{a}) = G_0(-\vek{b},\vek{a}-\vek{b}) = G_0(\vek{b},\vek{a})$, in consistency with (\ref{eq:sym0}).

\section{Inverse relations}
So far we have obtained the expressions of the four $\gamma$3PCFs as functions of the $\kappa$3PCFs. Written in a short form, they are
\eq{
\label{eq:conclu_relation}
\Gamma^{(i)}_{\rm cart} =  -\frac{1}{\pi^3} G_i * K^{(i)}\,, 
}
where $i$ runs from 0 to 3. The forms of the $G_i$ kernels are given by (\ref{eq:G2p}), (\ref{eq:G12p}), (\ref{eq:G22p}), and (\ref{eq:G32p}). 

These relations can be inverted. We define the kernels of the inverse relations to be $G'_i$, i.e.
\eq{
\label{eq:inv_relation}
K^{(i)} =  -\frac{1}{\pi^3} G'_i * \Gamma^{(i)}_{\rm cart}\,. 
}
Using the convolution theorem, it is apparent from (\ref{eq:conclu_relation}) and (\ref{eq:inv_relation}) that
\eq{
\label{eq:GGp}
\br{-\frac{1}{\pi^3} \tilde{G}_i} \cdot \br{-\frac{1}{\pi^3} \tilde{G}'_i} = 1\,.
}
From the corresponding Fourier plane relations of (\ref{eq:conclu_relation}), we also know 
\eq{
\label{eq:fGs}
-\frac{\tilde{G}_0}{\pi^3} = \expo{2\ic\br{\beta_1+\beta_2+\beta_3}}\,,\ \ 
-\frac{\tilde{G}_1}{\pi^3} = \expo{2\ic\br{-\beta_1+\beta_2+\beta_3}}\,,\ \ 
-\frac{\tilde{G}_2}{\pi^3} = \expo{2\ic\br{\beta_1-\beta_2+\beta_3}}\,,\ \ 
-\frac{\tilde{G}_3}{\pi^3} = \expo{2\ic\br{\beta_1+\beta_2-\beta_3}}\,,
}
which implies
\eq{
\label{eq:GGstarnorm}
\tilde{G}_0 \tilde{G}_0^* = \tilde{G}_1 \tilde{G}_1^* = \tilde{G}_2 \tilde{G}_2^* = \tilde{G}_3 \tilde{G}_3^* = \pi^6\,.
}

Comparing (\ref{eq:GGp}) and (\ref{eq:GGstarnorm}) one has
\eq{
 \tilde{G}'_i = \tilde{G}_i^* \,,
}
and further,
\eq{
 G'_i = G_i^* \,,
}
i.e. the convolution kernel for the inverse relation is the complex conjugate of the original kernel.

This property of the convolution kernel has its root in the fact that $\tilde{\gamma}$ and $\tilde{\kappa}$ differ only by a phase factor. This fact also endows the convolution kernels in the 1-pt and 2-pt relations between $\gamma$ and $\kappa$ with the same property. As is well known for the 1-pt relation, the inverse relation of (\ref{eq:KS}) is 
\eq{
\label{eq:invKS}
\kappa(\vek{x}) = \frac{1}{\pi} \int \dd^2 y\; \gamma(\vek{y}) \; \mathcal{D}^*(\vek{x}-\vek{y})\,,
}
where the kernel is the complex conjugate of the K-S kernel $\cal{D}$. The inverse relation of the 2-pt relation (\ref{eq:defineF}) can also be shown to be
\eq{
\label{eq:invdefF}
\langle \kappa\kappa \rangle (\vek{x}) =  \frac{1}{\pi^2}\int \dd^2 y \; \langle \gamma\gamma \rangle (\vek{y}) \; F^*(\vek{x}-\vek{y}) \,.
}

\section{Condition of 3pt E/B decomposition}
\label{sec:EB}
Being mathematically a polarization field, the shear field can be decomposed into a curl-free component and a divergence-free component, usually called the E-mode and the B-mode, respectively. Performing such a decomposition when treating cosmic shear data has long been recognized as a necessity, since it provides a valuable check on the possible systematics \citep[e.g.][]{crittenden02,pen02}. 

The E/B-mode decomposition can be done either on the shear field itself \citep[e.g.][]{bunn03, bunn10}, or at the level of correlation functions \citep[e.g.][]{s06}. The complex survey geometry after masking, which is especially characteristic for a lensing survey \citep[e.g.][]{erben09}, renders the first option barely feasible, and singles out the correlation function as the basic statistic to be applied directly to the data. Thus the natural way to perform the E/B-mode decomposition on cosmic shear data is to derive statistics based on the shear correlation functions. 

A commonly used statistic for this purpose is the aperture mass statistic which can be expressed as a linear combination of $\xi_+$ and $\xi_-$ \citep{s02},
\eqs{
\label{eq:mapxi}
\ba{M_{\rm ap}^2}(\theta)  =& \frac{1}{2} \int_0^{\infty} \frac{\dd\vartheta \; \vartheta}{\theta^2} \bb{\xi_+(\vartheta) T^{\rm ap}_+\br{\frac{\vartheta}{\theta}} + \xi_-(\vartheta) T^{\rm ap}_-\br{\frac{\vartheta}{\theta}}}\,,\\
\ba{M_{\perp}^2}(\theta)  =& \frac{1}{2} \int_0^{\infty} \frac{\dd\vartheta \; \vartheta}{\theta^2} \bb{\xi_+(\vartheta) T^{\rm ap}_+\br{\frac{\vartheta}{\theta}} - \xi_-(\vartheta) T^{\rm ap}_-\br{\frac{\vartheta}{\theta}}}\,,
}
where the forms of the weight functions $T^{\rm ap}_+$ and $T^{\rm ap}_-$ are given explicitly in \citet{s02}. The chosen forms of the weight functions guarantee that $\ba{M_{\rm ap}^2}$ responds only to the E-mode and $\ba{M_{\perp}^2}$ only to the B-mode.

The aperture mass statistics has been generalized to 3-pt level by \citet{jarvis04} and \citet{kilbinger05}, and is the only statistics available up to now which allows an E/B-mode decomposition at the 3-pt level. However, as found by \citet{kilbinger06}, it cannot ensure a clean E/B-mode decomposition when applied to real data. The lack of shear-correlation measurements on small and large scales, which arises from the inability of shape measurement for close projected galaxy pairs and the finite field size, prohibits one from performing the integral in (\ref{eq:mapxi}) from zero to infinity, and thus introduces a mixing of the E- and B-modes.

In recent years, there have been several efforts to construct better statistics which allow E/B-mode decomposition \citep{SK07,eifler10,fu10,s10}, all of them focusing on the cosmic shear 2-pt statistics. These new statistics are based on the idea that the weight functions $T^{\rm ap}_+$ and $T^{\rm ap}_-$ used in the aperture mass statistics are just one example out of the many possibilities. In general one can define second-order statistics in the form \citep{SK07}
\eqs{
\label{eq:EEandBB}
\textrm{EE} &= \int_0^{\infty} \vartheta\; \dd \vartheta\;\bb{ \xi_+(\vartheta) T_+(\vartheta) + \xi_-(\vartheta) T_-(\vartheta) }\,,\\
\textrm{BB} &= \int_0^{\infty} \vartheta\; \dd \vartheta\;\bb{ \xi_+(\vartheta) T_+(\vartheta) - \xi_-(\vartheta) T_-(\vartheta) }\,,
}
for which the condition that $\textrm{EE}$ responds only to E-mode and $\textrm{BB}$ only to B-mode is found to be
\eqs{
\label{eq:2pEBcondition}
\int_0^{\infty} & \vartheta\; \dd \vartheta\; T_+(\vartheta) J_0(\ell \vartheta) = \int_0^{\infty} \vartheta\; \dd \vartheta\; T_-(\vartheta) J_4(\ell \vartheta)\,, \ \rm{or\ equivalently}\\
&T_+(\vartheta) = T_-(\vartheta) + \int_{\vartheta}^{\infty} \theta\; \dd \theta\; T_-(\theta) \br{\frac{4}{\theta^2} - \frac{12\vartheta^2}{\theta^4}}\,.
}
Note that instead of being functions of the separation length as the aperture mass statistics, $\textrm{EE}$ and $\textrm{BB}$ are just numbers. At first sight one seems to have reduced the information quantity by integrating over the scale dependence in (\ref{eq:EEandBB}). In fact, the information can be easily regained by constructing a set of weight functions satisfying (\ref{eq:2pEBcondition}). As one example, $\ba{M_{\rm ap}^2}(\theta)$ and $\ba{M_{\perp}^2}(\theta)$ for any $\theta$ value can be reconstructed in the framework of (\ref{eq:EEandBB}) by specifying $T_+(\vartheta) = T^{\rm ap}_+\br{{\vartheta}/{\theta}}/{\theta^2}$ and $T_-(\vartheta) = T^{\rm ap}_-\br{{\vartheta}/{\theta}}/{\theta^2}$.

Since the condition for E/B-mode decomposition (\ref{eq:2pEBcondition}) still leaves large freedom for the choice of the weight functions, one can construct statistics which fulfill additional constraints, e.g. a finite support over the separation length. If one requires that $\textrm{EE}$ and $\textrm{BB}$ respond only to $\xi_+(\vartheta)$ and $\xi_-(\vartheta)$ with $ \vartheta_{\rm min}<\vartheta <\vartheta_{\rm max}$, where $\vartheta_{\rm min}$ and $\vartheta_{\rm max}$ are the chosen small- and large-scale cutoff, $T_+(\vartheta)$ and $T_-(\vartheta)$ must vanish outside the same range. Since $T_+$ and $T_-$ are interrelated by (\ref{eq:2pEBcondition}), one can specify only one of them to satisfy this constraint. The requirement that the other weight function also vanishes outside the specified range needs to be put as additional integral constraints. As shown by \citet{SK07}, if one chooses $T_-$ to vanish for $\vartheta < \vartheta_{\rm min}$ and $\vartheta > \vartheta_{\rm max}$, then to allow an E/B-mode decomposition on a finite interval $\vartheta_{\rm min} < \vartheta < \vartheta_{\rm max}$, $T_-$ has to satisfy additionally,  
\eq{
\label{eq:2pEBconditionfinite}
\int_{\vartheta_{\rm min}}^{\vartheta_{\rm max}}  \frac{\dd \vartheta}{\vartheta}\; T_-(\vartheta) = 0 = \int_{\vartheta_{\rm min}}^{\vartheta_{\rm max}}  \frac{\dd \vartheta}{\vartheta^3}\; T_-(\vartheta)\,,
}
which would guarantee that $T_+$ vanishes for $\vartheta < \vartheta_{\rm min}$ and $\vartheta > \vartheta_{\rm max}$.

Similar statistics are needed at the 3-pt level as well. The first step required is to formulate the conditions for 3-pt weight functions to allow E/B-mode decomposition, in analogy to (\ref{eq:2pEBcondition}). As we will show in this section, the relations between the $\gamma$3PCFs and $\kappa$3PCFs that we derived provide a natural way of formulating such conditions. 

A pure E-mode shear 3-pt statistics is related only to the E-mode $\kappa$3PCF $\ba{\kappa^{\rm{E}}\kappa^{\rm{E}}\kappa^{\rm{E}}}$ but not to other 3PCFs with $\kappa^{\rm{B}}$ contribution. Therefore we first write the 3PCFs of $\kappa^{\rm{E}}$ and $\kappa^{\rm{B}}$ as linear combinations of the real and imaginary parts of the $K^{(i)}$'s, using (\ref{eq:KEB}), and then relate them with the $\Gamma_{\rm cart}$'s through (\ref{eq:inv_relation}), as the $\Gamma_{\rm cart}$'s are the directly measurable statistics from a lensing survey. The results read
\eqs{
\label{eq:kappaEB2gamma}
\langle \kappa^{\rm{E}}\kappa^{\rm{E}}\kappa^{\rm{E}} \rangle =\  & \frac{1}{4} \textit{Re}\bb{\ K^{(0)} + K^{(1)} + K^{(2)} + K^{(3)}\ } 
 = -\frac{1}{4\pi^3} \textit{Re}\bb{\ G_0^* * \Gamma^{(0)}_{\rm cart} +  G_1^* * \Gamma^{(1)}_{\rm cart} + G_2^* * \Gamma^{(2)}_{\rm cart} + G_3^* * \Gamma^{(3)}_{\rm cart}\ } \,,\\
\langle \kappa^{\rm{E}}\kappa^{\rm{B}}\kappa^{\rm{B}} \rangle =\  & \frac{1}{4} \textit{Re}\bb{-K^{(0)} - K^{(1)} + K^{(2)} + K^{(3)}} 
 = -\frac{1}{4\pi^3} \textit{Re}\bb{- G_0^* * \Gamma^{(0)}_{\rm cart} -  G_1^* * \Gamma^{(1)}_{\rm cart} + G_2^* * \Gamma^{(2)}_{\rm cart} + G_3^* * \Gamma^{(3)}_{\rm cart}} \,,\\
\langle \kappa^{\rm{B}}\kappa^{\rm{E}}\kappa^{\rm{B}} \rangle =\  & \frac{1}{4} \textit{Re}\bb{-K^{(0)} + K^{(1)} - K^{(2)} + K^{(3)}} 
 = -\frac{1}{4\pi^3} \textit{Re}\bb{- G_0^* * \Gamma^{(0)}_{\rm cart} +  G_1^* * \Gamma^{(1)}_{\rm cart} - G_2^* * \Gamma^{(2)}_{\rm cart} + G_3^* * \Gamma^{(3)}_{\rm cart}} \,,\\
\langle \kappa^{\rm{B}}\kappa^{\rm{B}}\kappa^{\rm{E}} \rangle =\  & \frac{1}{4} \textit{Re}\bb{-K^{(0)} + K^{(1)} + K^{(2)} - K^{(3)}} 
 = -\frac{1}{4\pi^3} \textit{Re}\bb{- G_0^* * \Gamma^{(0)}_{\rm cart} +  G_1^* * \Gamma^{(1)}_{\rm cart} + G_2^* * \Gamma^{(2)}_{\rm cart} - G_3^* * \Gamma^{(3)}_{\rm cart}} \,,\\
}
and
\eqs{
\langle \kappa^{\rm{B}}\kappa^{\rm{E}}\kappa^{\rm{E}} \rangle =\  & \frac{1}{4} \textit{Im}\bb{\ K^{(0)} - K^{(1)} + K^{(2)} + K^{(3)}\ } 
 = -\frac{1}{4\pi^3} \textit{Im}\bb{\ G_0^* * \Gamma^{(0)}_{\rm cart} -  G_1^* * \Gamma^{(1)}_{\rm cart} + G_2^* * \Gamma^{(2)}_{\rm cart} +  G_3^* * \Gamma^{(3)}_{\rm cart}\ } \,,\\
\langle \kappa^{\rm{E}}\kappa^{\rm{B}}\kappa^{\rm{E}} \rangle =\  & \frac{1}{4} \textit{Im}\bb{\ K^{(0)} + K^{(1)} - K^{(2)} + K^{(3)}\ }
 = -\frac{1}{4\pi^3} \textit{Im}\bb{\ G_0^* * \Gamma^{(0)}_{\rm cart} +  G_1^* * \Gamma^{(1)}_{\rm cart} -  G_2^* * \Gamma^{(2)}_{\rm cart} +  G_3^* * \Gamma^{(3)}_{\rm cart}\ } \,,\\
\langle \kappa^{\rm{E}}\kappa^{\rm{E}}\kappa^{\rm{B}} \rangle =\  & \frac{1}{4} \textit{Im}\bb{\ K^{(0)} + K^{(1)} + K^{(2)} - K^{(3)}\ }
 = -\frac{1}{4\pi^3} \textit{Im}\bb{\ G_0^* * \Gamma^{(0)}_{\rm cart} +  G_1^* * \Gamma^{(1)}_{\rm cart} + G_2^* * \Gamma^{(2)}_{\rm cart} -  G_3^* * \Gamma^{(3)}_{\rm cart}\ } \,,\\
\langle \kappa^{\rm{B}}\kappa^{\rm{B}}\kappa^{\rm{B}} \rangle =\  & \frac{1}{4} \textit{Im}\bb{-K^{(0)} + K^{(1)} + K^{(2)} + K^{(3)}} 
 = -\frac{1}{4\pi^3} \textit{Im}\bb{ - G_0^* * \Gamma^{(0)}_{\rm cart} +  G_1^* * \Gamma^{(1)}_{\rm cart} + G_2^* * \Gamma^{(2)}_{\rm cart} + G_3^* * \Gamma^{(3)}_{\rm cart}} \,,\\
}
which shows how the E- and B-mode $\kappa$3PCFs can be computed when the full information of the $\Gamma_{\rm cart}$'s is available. In the ideal case that there exists no noise or systematical effects, only the E-mode term $\kappa$3PCF $\ba{\kappa^{\rm{E}}\kappa^{\rm{E}}\kappa^{\rm{E}}}$ is expected to be non-zero, since it corresponds to the correlation in the physical density field which leads to the correlation in the shear signal. 

Following the ideas of \citet{SK07}, we construct a new statistic
\eq{
\label{eq:EEEkappa}
\textrm{EEE} = \int \dd^2 x_1 \int \dd^2 x_2 \;\ba{\kappa^{\rm{E}}\kappa^{\rm{E}}\kappa^{\rm{E}}}(\vek{x_1},\vek{x_2})\; U(\vek{x_1},\vek{x_2})\,,
}
which by definition responds only to E-mode. With the help of (\ref{eq:kappaEB2gamma}) we can link $\textrm{EEE}$ to the observable $\Gamma_{\rm cart}$'s, as
\eqs{
\label{eq:EEEgamma}
\textrm{EEE} =& -\frac{1}{4\pi^3} \int \dd^2 x_1 \int \dd^2 x_2 \; U(\vek{x_1},\vek{x_2})\; \textit{Re} \bb{ \int \dd^2 y_1 \int \dd^2 y_2 \;  \sum_{i=0}^3 \ G_i^*(\vek{x_1}-\vek{y_1},\vek{x_2}-\vek{y_2})\; \Gamma^{(i)}_{\rm cart}(\vek{y_1},\vek{y_2})} \\
= &   -\frac{1}{4\pi^3}  \textit{Re} \bb{ \int \dd^2 y_1 \int \dd^2 y_2 \;  \sum_{i=0}^3 \Gamma^{(i)}_{\rm cart}(\vek{y_1},\vek{y_2}) \int \dd^2 x_1 \int \dd^2 x_2 \; U(\vek{x_1},\vek{x_2})\;  G_i^*(\vek{x_1}-\vek{y_1},\vek{x_2}-\vek{y_2}) } \\
= &  -\frac{1}{4\pi^3}\textit{Re} \bb{ \int \dd^2 y_1 \int \dd^2 y_2 \;  \sum_{i=0}^3 \Gamma^{(i)}_{\rm cart}(\vek{y_1},\vek{y_2}) \br{G_i^* * U}(\vek{y_1},\vek{y_2}) }\,,
}
where in the first equation we have specified $U$ to be a real function, and in the second equation we have used the fact that $G_i(-\vek{a},-\vek{b}) = G_i(\vek{a}, \vek{b})$. 

Denoting 
\eq{
G_i^* * U=:T^{(i)}\,,
}
the expression of EEE (\ref{eq:EEEgamma}) has a similar form as (\ref{eq:EEandBB}). We can see in this form that $\textrm{EEE}$ responds only to the E-mode if the weight function $T$'s satisfy
\eq{
\label{eq:conditionEEE}
T^{(0)}*G_0 = U = T^{(1)}*G_1 = T^{(2)}*G_2 = T^{(3)}*G_3\,.
}
with $U$ being a real function. One can easily verify that these conditions are also necessary conditions. Noticing that $T^{(i)}*G_i$ is the corresponding weight on $K^{(i)}$, the condition that these functions being purely real is required to separate the parity-violating and non-violating terms in (\ref{eq:KEB}). In addition, (\ref{eq:conditionEEE}) is required to cancel the parity non-violating B-mode terms $\langle \kappa^{\rm{E}}\kappa^{\rm{B}}\kappa^{\rm{B}} \rangle$, $\langle \kappa^{\rm{B}}\kappa^{\rm{E}}\kappa^{\rm{B}} \rangle$ and $\langle \kappa^{\rm{B}}\kappa^{\rm{B}}\kappa^{\rm{E}} \rangle$.

The statistics containing the contribution from only one of the B-mode terms can be constructed in the same way. Omitting the arguments for notational simplicity, they can be expressed as
\eqs{
\label{eq:Bmodecontri}
\textrm{EBB} =& \frac{1}{4}\textit{Re}\bb{\iint  \sum_{i=0}^3 T^{(i)} \;\Gamma^{(i)}_{\rm cart}} \ \  \textrm{with}\ \  T^{(0)}*G_0 = T^{(1)}*G_1 = -T^{(2)}*G_2 = -T^{(3)}*G_3 \,,\\
\textrm{BEB} =& \frac{1}{4}\textit{Re}\bb{\iint  \sum_{i=0}^3 T^{(i)} \;\Gamma^{(i)}_{\rm cart}} \ \  \textrm{with}\ \  T^{(0)}*G_0 = -T^{(1)}*G_1 = T^{(2)}*G_2 = -T^{(3)}*G_3 \,,\\
\textrm{BBE} =& \frac{1}{4}\textit{Re}\bb{\iint  \sum_{i=0}^3 T^{(i)} \;\Gamma^{(i)}_{\rm cart}} \ \  \textrm{with}\ \  T^{(0)}*G_0 = -T^{(1)}*G_1 = -T^{(2)}*G_2 = T^{(3)}*G_3 \,,\\
} 
and
\eqs{
\label{eq:parityvio}
\textrm{BEE} =& \frac{1}{4}\textit{Im}\bb{\iint  \sum_{i=0}^3 T^{(i)} \;\Gamma^{(i)}_{\rm cart}} \ \  \textrm{with}\ \   T^{(0)}*G_0 = -T^{(1)}*G_1 = T^{(2)}*G_2 = T^{(3)}*G_3 \,,\\
\textrm{EBE} =& \frac{1}{4}\textit{Im}\bb{\iint  \sum_{i=0}^3 T^{(i)} \;\Gamma^{(i)}_{\rm cart}} \ \  \textrm{with}\ \   T^{(0)}*G_0 = T^{(1)}*G_1 = -T^{(2)}*G_2 = T^{(3)}*G_3 \,,\\
\textrm{EEB} =& \frac{1}{4}\textit{Im}\bb{\iint  \sum_{i=0}^3 T^{(i)} \;\Gamma^{(i)}_{\rm cart}} \ \  \textrm{with}\ \   T^{(0)}*G_0 = T^{(1)}*G_1 = T^{(2)}*G_2 = -T^{(3)}*G_3 \,,\\
\textrm{BBB} =& \frac{1}{4}\textit{Im}\bb{\iint  \sum_{i=0}^3 T^{(i)} \;\Gamma^{(i)}_{\rm cart}} \ \  \textrm{with}\ \   T^{(0)}*G_0 = -T^{(1)}*G_1 = -T^{(2)}*G_2 = -T^{(3)}*G_3 \,.\\
}
For all of them the condition that $T^{(0)}*G_0$ is real has been imposed.

The four parity violating statistics (\ref{eq:parityvio}) can be used as a check on parity violating systematical errors, while the other B-mode statistics (\ref{eq:Bmodecontri}) allows for a further examination of the B-modes. 

With (\ref{eq:GGstarnorm}) one can easily invert the conditions on the weight functions to express the weight function $T$'s directly in terms of each other. Take the conditions for \textrm{EEE} (\ref{eq:conditionEEE}) for example. One can write the Fourier transforms of $T^{(1)}$, $T^{(2)}$ and $T^{(3)}$ as functions of the Fourier transform of $T^{(0)}$ as
\eq{
\label{eq:T0Tirelation}
\tilde{T}^{(1)} = \frac{1}{\pi^6} \tilde{T}^{(0)}\;\tilde{G}_0\tilde{G}_1^*\,,\ \ \tilde{T}^{(2)} = \frac{1}{\pi^6} \tilde{T}^{(0)}\;\tilde{G}_0\tilde{G}_2^*\,,\ \ \tilde{T}^{(3)} = \frac{1}{\pi^6} \tilde{T}^{(0)}\;\tilde{G}_0\tilde{G}_3^*\,. 
}
In order to simplify these relations, we now attempt to give simple expressions for $\tilde{G}_0\tilde{G}_i^*$ for $i=1,2,3$. With (\ref{eq:fGs}) one has $\tilde{G}_0(\vek{\ell_1},\vek{\ell_2}) \tilde{G}_1^*(\vek{\ell_1},\vek{\ell_2}) = \pi^6 \expo{4\ic \beta_1}$, which does not depend on $\vek{\ell_2}$. Noticing that $\tilde{F} = \pi^2 \expo{4\ic \beta}$ (see Sect.\;\ref{sec:fouriertransfs}), we actually have $\tilde{G}_0(\vek{\ell_1},\vek{\ell_2}) \tilde{G}_1^*(\vek{\ell_1},\vek{\ell_2}) = \pi^4 \tilde{F}(\vek{\ell_1})$, which yields in configuration space
\eq{
\label{eq:G01}
\br{G_0*G_1^*}(\vek{a},\vek{b}) = \pi^4 F(\vek{a})\; \delta^{(2)}_{\rm D}(\vek{b}) \,.
}
Similarly one can derive that
\eq{
\label{eq:G023}
\br{G_0*G_2^*}(\vek{a},\vek{b}) = \pi^4 F(\vek{b})\; \delta^{(2)}_{\rm D}(\vek{a})\,,\ \ \br{G_0*G_3^*}(\vek{a},\vek{b}) = \pi^4 F(\vek{a})\; \delta^{(2)}_{\rm D}(\vek{b}-\vek{a}) \,.
}
Inserting (\ref{eq:G01}) and (\ref{eq:G023}) into (\ref{eq:T0Tirelation}), one obtains
\eq{
\label{eq:T01}
T^{(1)}(\vek{x_1},\vek{x_2}) = \frac{1}{\pi^2} \int \dd^2 y \; T^{(0)}(\vek{x_1}-\vek{y},\vek{x_2}) F(\vek{y}) =  \frac{2}{\pi} \int \dd^2 y \; T^{(0)}(\vek{x_1}-\vek{y},\vek{x_2}) \frac{\vek{y}}{\vek{y}^{*3}}\,,
} 
\eq{
\label{eq:T02}
T^{(2)}(\vek{x_1},\vek{x_2}) = \frac{1}{\pi^2} \int \dd^2 y \; T^{(0)}(\vek{x_1},\vek{x_2}-\vek{y}) F(\vek{y}) = \frac{2}{\pi} \int \dd^2 y \; T^{(0)}(\vek{x_1},\vek{x_2}-\vek{y}) \frac{\vek{y}}{\vek{y}^{*3}}\,,
} 
\eq{
\label{eq:T03}
T^{(3)}(\vek{x_1},\vek{x_2}) = \frac{1}{\pi^2} \int \dd^2 y \; T^{(0)}(\vek{x_1}-\vek{y},\vek{x_2}-\vek{y}) F(\vek{y})=\frac{2}{\pi} \int \dd^2 y \; T^{(0)}(\vek{x_1}-\vek{y},\vek{x_2}-\vek{y}) \frac{\vek{y}}{\vek{y}^{*3}}\,,
} 
where we have used the expression of $F$ (\ref{eq:Fform}).

To summarize, if one chooses an arbitrary form of $T^{(0)}$ which makes $T^{(0)}*G_0$ real, constructs $T^{(1)}$, $T^{(2)}$ and $T^{(3)}$ according to ($\ref{eq:T01}$), ($\ref{eq:T02}$) and ($\ref{eq:T03}$), and uses these weight functions to weight the measured $\Gamma^{(i)}_{\rm cart}$'s, then the resulting statistic \textrm{EEE} as defined in (\ref{eq:EEEgamma}) receives contributions only from $\langle \kappa^{\rm{E}}\kappa^{\rm{E}}\kappa^{\rm{E}}\rangle$ but not the terms affected by the B-mode. The B-mode statistics can be obtained from the measured $\Gamma^{(i)}_{\rm cart}$'s through a similar procedure. Equations ($\ref{eq:T01}$)-($\ref{eq:T03}$) are the analogue of (\ref{eq:2pEBcondition}) for 3-pt functions.

We note again that $\Gamma^{(1)}_{\rm cart}$, $\Gamma^{(2)}_{\rm cart}$ and $\Gamma^{(3)}_{\rm cart}$ are not independent under transformation of their arguments. Thus the statistics \textrm{EEE} (\ref{eq:EEEgamma}) as well as the B-mode statistics (\ref{eq:Bmodecontri}) and (\ref{eq:parityvio}) can all be written in terms of linear combinations of $\Gamma^{(0)}_{\rm cart}$ and $\Gamma^{(1)}_{\rm cart}$ alone. However we shall keep the current redundancy since it allows for simple analytical expressions of the relations between the weight functions. 

So far one still has much freedom in choosing the form of $T^{(0)}(\vek{x_1},\vek{x_2})$. This freedom can be exploited to construct statistics which do not respond to the $\gamma$3PCF at smaller or larger angular separations than can be probed by the survey. We leave this to future work. 

\section{Numerical evaluation}

\subsection{Design of the sampling grid}

We have written configuration space relations between weak lensing statistics in the form of convolutions, e.g. (\ref{eq:Frelation}) and (\ref{eq:Grelation}), where the convolution kernels are complex, have non-trivial spin numbers, and feature singularities. The convolutions can be performed numerically, but special care must be taken of these properties of the integration kernel.  

\begin{figure}[h]
\centering
\includegraphics[width=16cm]{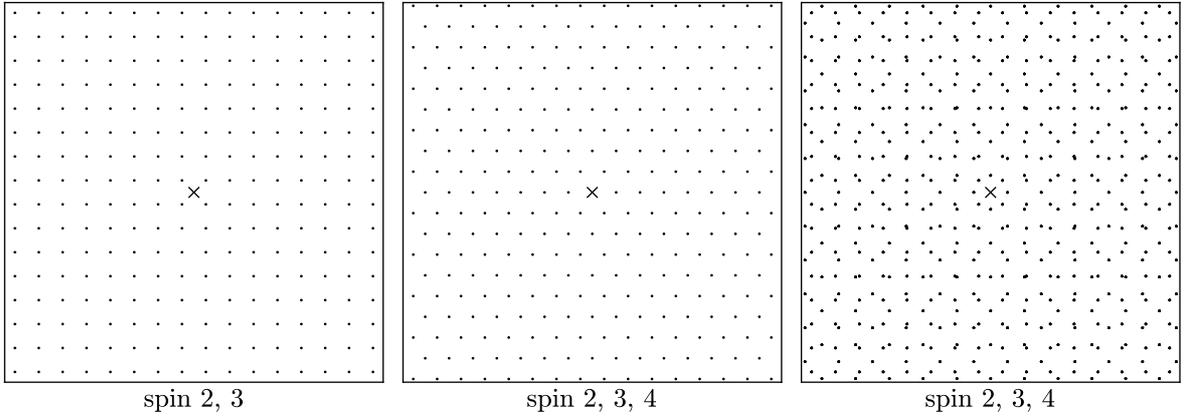}
\caption{ Examples of sampling grids applicable for 2D integration over singular kernels of different spin values. The cross in the center indicates the position of the singularity in the integration kernel.}
\label{fig:grids}
\end{figure}

One can take the K-S kernel $-1/\vek{z}^{*2}$ as an example of this kind of convolution kernel. The K-S kernel has an integer spin of 2, so an azimuthal integration of the kernel around its singularity at $z=0$ should give zero, i.e. the values of the kernel along the circle cancel themselves due to their opposite phases. This property renders its singularity harmless, but entails the condition that the sampling grid should guarantee the cancellation. Such a requirement of a special grid design has already been realized in early lensing mass reconstruction works \citep[e.g.][]{seitz96}. For a spin-2 kernel like the K-S kernel, a common square grid already suffices if the singularity is placed at a center of rotational symmetry, i.e. either onto a grid point or at the center of four grid points. In the case of the former, the grid point at the singularity has to be discarded. The latter, as shown by the left panel in Fig.\,\ref{fig:grids}, is a better choice considering the sampling homogeneity. 

\begin{figure}[h]
\begin{minipage}[c]{.6\textwidth}
\hspace{1cm}
\includegraphics[width=9cm]{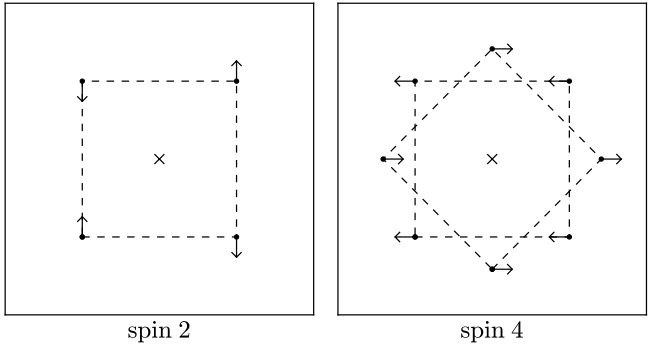}
\end{minipage}
\begin{minipage}[c]{.35\textwidth}
\caption{A square grid guarantees phase cancellation around the singularity of spin-2 kernels (left panel), but not that of spin-4 kernels (right panel). The crosses indicate the position of the singularity of the integration kernel, while the dots are the grid points closest to the singularity. The polar angles of the kernel at the grid points are indicated by the directions of the arrows. For a spin-2 kernel they cancel each other on a square grid already. For a spin-4 kernel they cancel each other if the square grid is duplicated, rotated 45 degrees and put on top of the original grid. }
\hspace{1cm}
\label{fig:spin2spin4}
\end{minipage}
\end{figure}

In general we need to deal with convolution kernels with different spin numbers. For example, the kernel $F$ between $\gamma$2PCF and $\kappa$2PCF (\ref{eq:Fform}), which is proportional to $\vek{z}/\vek{z^{*3}}$, has a spin of 4. In this case the square grid cannot guarantee the phase cancellation around the singularity any more (see Fig.\;\ref{fig:spin2spin4}). With a similar analysis as shown in Fig.\,\ref{fig:spin2spin4}, one can see that a triangular grid (middle panel, Fig.\,\ref{fig:grids}) can actually guarantee the phase cancellation around the singularity of a spin-4 kernel. When using a triangular grid, the singularity can also be put either on a grid point or at the center of three grid points. The former choice loses the grid point at the singularity, but is applicable to spin-3 kernels where the latter fails. 

To achieve a high numerical accuracy, it is required that the circle integral around the singularity is well-sampled. If one uses a square (triangular) grid, the innermost circle is only sampled by four (six) grid points, which is not enough for many $\kappa$-models. To remedy this, one can duplicate the sampling grid, rotate it around the singularity, and put it on top of the original grid. We show the result with the square grid and 45 degrees of rotation in the right panel of Fig.\,\ref{fig:grids}. The resulting grid is also applicable for spin-4 kernels, as shown in the right panel of Fig.\;\ref{fig:spin2spin4}. This non-standard way of constructing sampling grids, although not creating the best grids in terms of sampling efficiency, deals very well with the singularity of the integration kernel, and can easily generate sampling grids applicable for kernels of any spin number. We use this kind of grid in our numerical sampling.
 
Additional complications arise when performing the integration for 3-pt statistics. There are two 2D integrals in (\ref{eq:defineG}), (\ref{eq:defineG1}), (\ref{eq:defineG2}) and (\ref{eq:defineG3}). The corresponding integration kernels (\ref{eq:G2p}), (\ref{eq:G12p}), (\ref{eq:G22p}) and (\ref{eq:G32p}) all have three singularities. Luckily we can split each integration kernel into four additive terms and perform the integrals over each of them separately. Moreover, one can apply translational shifts to the integrands so that in each 2D integral there is only one singularity in the integration kernel. The singularity can also be shifted to the origin of the grids for numerical simplicity. After all these procedures, the four relations can be written as
\eqs{
\label{eq:Grelationterm}
\Gamma^{(0)}_{\rm cart}(\vek{x_1},\vek{x_2}) 
= & \frac{2}{\pi^2}\;\int \dd^2 y_1 \int \dd^2 y_2\; \Bigg\{ - \bb{\langle \kappa\kappa\kappa \rangle (\vek{y_2}+\vek{x_1},\vek{y_1}+\vek{y_2}+\vek{x_2})+\langle \kappa\kappa\kappa \rangle (\vek{y_1}+\vek{x_1}+\vek{y_2},\vek{y_2}+\vek{x_2}) }\; \frac{1}{\vek{y_1}^{*2}} \; \frac{\vek{y_2}}{\vek{y_2}^{*3}}\\
& + \bb{\langle \kappa\kappa\kappa \rangle (\vek{y_2}+\vek{x_1},\vek{y_1}+\vek{y_2}+\vek{x_2})+\langle \kappa\kappa\kappa \rangle (\vek{y_1}+\vek{x_1}+\vek{y_2},\vek{y_2}+\vek{x_2}) }\; \frac{1}{\vek{y_1}^{*3}} \; \frac{\vek{y_2}}{\vek{y_2}^{*2}}  \Bigg\} \,,
}
\eqs{
\label{eq:G1relation}
\Gamma^{(1)}_{\rm cart}(\vek{x_1},\vek{x_2}) 
=& -\frac{1}{\pi} \int \dd^2 y\; \br{\langle \kappa^*\kappa\kappa \rangle(\vek{x_1},\vek{y}+\vek{x_2})+\langle \kappa^*\kappa\kappa \rangle(\vek{y}+\vek{x_1},\vek{y}+\vek{x_2})}\frac{1}{\vek{y}^{*2}} \\
&+ \frac{2}{\pi^2} \;\int \dd^2 y_1 \int \dd^2 y_2\; \br{\langle \kappa^*\kappa\kappa \rangle(\vek{y_2}+\vek{x_1},\vek{y_1}+\vek{x_2})-\langle \kappa^*\kappa\kappa \rangle(\vek{y_1}+\vek{y_2}+\vek{x_1},\vek{y_1}+\vek{x_2})} \frac{1}{\vek{y_1}^{*3}}\frac{1}{\vek{y_2}}\,,
}
\eqs{
\label{eq:G2relation}
\Gamma^{(2)}_{\rm cart}(\vek{x_1},\vek{x_2}) 
=& -\frac{1}{\pi} \int \dd^2 y\; \br{\langle \kappa\kappa^*\kappa \rangle(\vek{y}+\vek{x_1},\vek{y}+\vek{x_2})+\langle \kappa\kappa^*\kappa \rangle(\vek{y}+\vek{x_1},\vek{x_2})}\frac{1}{\vek{y}^{*2}} \\
&- \frac{2}{\pi^2} \;\int \dd^2 y_1 \int \dd^2 y_2\; \br{\langle \kappa\kappa^*\kappa \rangle(\vek{y_1}+\vek{x_1},\vek{y_1}+\vek{y_2}+\vek{x_2})-\langle \kappa\kappa^*\kappa \rangle(\vek{y_1}+\vek{x_1},\vek{y_2}+\vek{x_2})} \frac{1}{\vek{y_1}^{*3}}\frac{1}{\vek{y_2}}\,, 
}
\eqs{
\label{eq:G3relation}
\Gamma^{(3)}_{\rm cart}(\vek{x_1},\vek{x_2}) 
=& -\frac{1}{\pi} \int \dd^2 y\; \br{\langle \kappa\kappa\kappa^* \rangle(\vek{x_1},\vek{y}+\vek{x_2})+\langle \kappa\kappa\kappa^* \rangle(\vek{y}+\vek{x_1},\vek{x_2})}\frac{1}{\vek{y}^{*2}} \\
&- \frac{2}{\pi^2} \;\int \dd^2 y_1 \int \dd^2 y_2\; \br{\langle \kappa\kappa\kappa^* \rangle(\vek{x_1}+\vek{y_2},\vek{y_1}+\vek{y_2}+\vek{x_2})+\langle \kappa\kappa\kappa^* \rangle(\vek{y_1}+\vek{y_2}+\vek{x_1},\vek{x_2}+\vek{y_2})} \frac{1}{\vek{y_1}^{*3}}\frac{1}{\vek{y_2}}\,. 
}
We can see that the integration kernels are either spin-2, spin-4, or spin-3. All sampling grids shown in Fig.\,\ref{fig:grids} are applicable to spin-3 kernels.

\subsection{Numerical results for 2PCF}
We now construct several toy models for 2-pt and 3-pt convergence and shear correlations, and use them to test the relations derived as well as the numerical sampling method.

In the 2-pt case, we build two models for the convergence correlation function: $\ba{\kappa\kappa}(\vek{r}) = 1/r$, and $\ba{\kappa\kappa}(\vek{r}) = \expo{-r^2}$. Using the well-established $\xi_+-\xi_-$ relation (\ref{eq:xi+xi-}) we can obtain the corresponding models for the shear correlation function: $\ba{\gamma\gamma}(\vek{r}) = \expo{4\ic\phi_r}/r$, and $\ba{\gamma\gamma}(\vek{r}) = \expo{4\ic\phi_r}\bb{(r^4+4r^2+6)\expo{-r^2}+2r^2-6}/r^4$. 

\begin{figure}[h]
\begin{minipage}[h]{.65\textwidth}
\includegraphics[width=12cm]{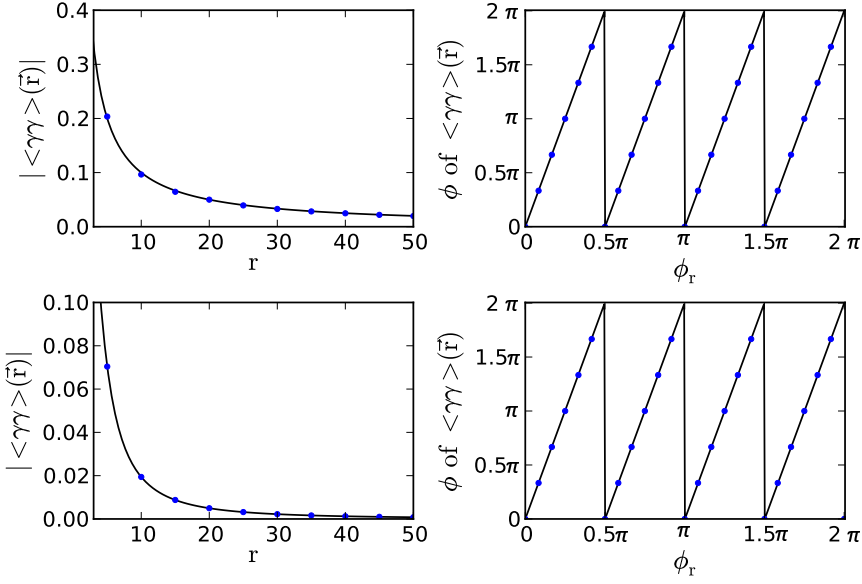}
\end{minipage}
\begin{minipage}[t]{.3\textwidth}
\vspace{-0.2cm}
\caption{ \textit{Left panels:} $\langle \gamma\gamma\rangle(\vek{r})$ with fixed $\phi_r = \pi/6$ as a function of $r$. \textit{Right panels:} Polar angle of $\langle \gamma\gamma\rangle(\vek{r})$ with fixed $r = 20$ as a function of $\phi_r$. The black curves are the expected values while the dots are the results from numerical evaluation of 2D integral in (\ref{eq:Frelation}). The $\langle \kappa\kappa\rangle$ used in the upper panels is $\langle \kappa\kappa\rangle(\vek{r}) = 1/r $ and in the lower panels $\langle \kappa\kappa\rangle(\vek{r}) = \textrm{exp}(-r^2)$.}
\label{fig:sample2p}
\end{minipage}
\end{figure}

Fig.\,\ref{fig:sample2p} shows the comparison between these shear correlation function models and the shear correlation functions sampled using (\ref{eq:Frelation}), with the corresponding convergence correlation function models as input. The numerically sampled values match the analytical models very well. 

\subsection{Numerical results for 3PCF}

\begin{figure}[h]
\begin{minipage}[c]{.7\textwidth}
\includegraphics[width=12.7cm]{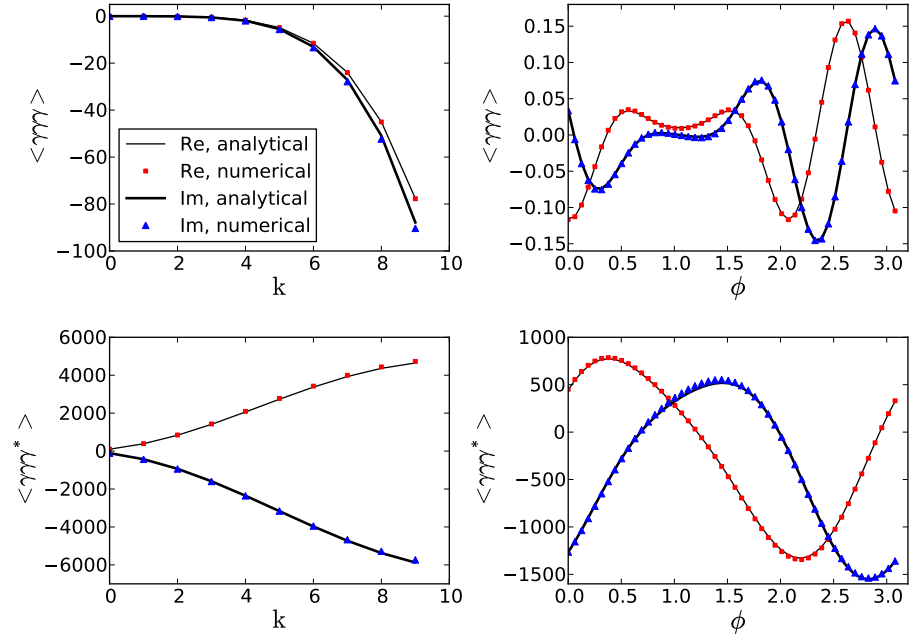}
\end{minipage}
\begin{minipage}[c]{.25\textwidth}
\vspace{3cm}
\caption{Comparison of numerically evaluated $\langle \gamma\gamma\gamma\rangle(\vek{x}, \vek{y})$ and $\langle \gamma\gamma\gamma^*\rangle(\vek{x}, \vek{y})$ with their analytical toy models described in this section adopting $\alpha = 0.05$. `Re' and `Im' indicate the real and imaginary parts of the shear correlation functions. \textit{Left panels:} The functions are evaluated at $\vek{x} = 0.25 k \expo{-\ic \pi /3}$ and $\vek{y}=0.17 k \expo{\ic \pi/8}$ for different values of $k$; \textit{Right panels:} The functions are evaluated at $\vek{x}=0.75 \expo{-\ic \pi/3}$ and $\vek{y}=0.45 \expo{\ic \phi}$ for 50 equally spaced $\phi$ values.}
\label{fig:3pmodel2}
\end{minipage}
\end{figure}

We build models for 3pt shear and convergence correlation functions via the 3pt correlation function of the deflection potential $\psi$. Suppose we evaluate the fields at positions $X$, $Y$ and $Z$. By definition we have

\eq{
\langle \kappa(\vek{X})\kappa(\vek{Y})\kappa(\vek{Z})\rangle = \br{\frac{1}{2}\nabla^2_X} \br{\frac{1}{2}\nabla^2_Y} \br{\frac{1}{2}\nabla^2_Z} \;  \langle \psi(\vek{X})\psi(\vek{Y})\psi(\vek{Z})\rangle\,,
}

\eq{
\langle \gamma(\vek{X})\gamma(\vek{Y})\gamma(\vek{Z})\rangle = \br{\frac{1}{2}\partial^2_X} \br{\frac{1}{2}\partial^2_Y} \br{\frac{1}{2}\partial^2_Z} \; \langle \psi(\vek{X})\psi(\vek{Y})\psi(\vek{Z})\rangle\,,
}
and
\eq{
\langle \gamma(\vek{X})\gamma(\vek{Y})\gamma^*(\vek{Z})\rangle = \br{\frac{1}{2}\partial^2_X} \br{\frac{1}{2}\partial^2_Y} \br{\frac{1}{2}\partial^{*2}_Z} \; \langle \psi(\vek{X})\psi(\vek{Y})\psi(\vek{Z})\rangle\,.
}

Now we assume a model for $ \langle \psi(\vek{X})\psi(\vek{Y})\psi(\vek{Z})\rangle$ as 
\eq{
\langle \psi(\vek{X})\psi(\vek{Y})\psi(\vek{Z})\rangle = \frac{1}{8 \alpha^6} \expo{-\alpha (x^2+y^2)}\,,
}
with $\vek{x} = \vek{X}-\vek{Z}$ and $\vek{y} = \vek{Y}-\vek{Z}$. This model is special in the sense that it does not depend on the angle between $\vek{x}$ and $\vek{y}$, but is nevertheless simple and rather local. The statistical homogeneity of the field enables us to write the 3pt correlation function as a function of two 2D spatial coordinates, which we have chosen here to be $\vek{x}$ and $\vek{y}$. 

Performing the derivatives, we find the corresponding 3pt shear and convergence correlation functions also depend only on $\vek{X}-\vek{Z} = \vek{x}$ and $\vek{Y}-\vek{Z} = \vek{y}$, and read
\eq{
\label{eq:3pk}
\langle \kappa\kappa\kappa\rangle (\vek{x}, \vek{y}) = 
 \Big[-4 +\alpha\br{6(x^2+y^2)+8 \vek{x}\cdot\vek{y}} - \alpha^2 \Big((x^2+y^2)^2  +4(x^2+y^2) \vek{x}\cdot\vek{y} +6 x^2 y^2 \Big)+\alpha^3 x^2 y^2 |\vek{x}+\vek{y}|^2 \Big]\; \frac{\expo{-\alpha (x^2+y^2)}}{\alpha^3} \,,
}

\eq{
\label{eq:3pg0}
\langle \gamma\gamma\gamma\rangle (\vek{x}, \vek{y}) = \vek{x}^2\vek{y}^2(\vek{x}+\vek{y})^2\expo{-\alpha (x^2+y^2)}\,,
}
and
\eq{
\label{eq:3pg1}
\langle \gamma\gamma\gamma^*\rangle (\vek{x}, \vek{y}) = \bb{2\vek{x}^2+8\vek{x}\vek{y}+2\vek{y}^2- 4\alpha \vek{x}\vek{y}\; |\vek{x}+\vek{y}|^2+\alpha^2 \vek{x}^2\vek{y}^2 (\vek{x}^*+\vek{y}^*)^2
}\expo{-\alpha (x^2+y^2)}\,.
}

Using (\ref{eq:3pk}) as the input model for $\langle \kappa\kappa\kappa\rangle$, we numerically evaluate $\langle \gamma\gamma\gamma\rangle(\vek{x}, \vek{y})$ and $\langle \gamma\gamma\gamma^*\rangle (\vek{x}, \vek{y})$ using (\ref{eq:Grelationterm}) and (\ref{eq:G3relation}). The results are then compared to the analytical models for the 3pt shear correlation functions (\ref{eq:3pg0}) and (\ref{eq:3pg1}). As shown in Fig.\;\ref{fig:3pmodel2}, the numerical evaluations closely match the analytical models.

Hence, we have proven numerically that the relations between $\gamma$3PCFs and $\kappa$3PCFs (\ref{eq:Grelationterm}) and (\ref{eq:G3relation}) are correct, no additional delta functions are needed. At the same time we have shown that these relations can be numerically evaluated with a high accuracy. Therefore these relations provide a better way of relating the observable shear signal to the underlying matter density field than the original way, i.e. using the relations between the $\gamma$3PCFs and the $\kappa$ bispectrum, since the latter does not allow for an easy accurate numerical evaluation.

\section{Conclusion}
We derived the relations between the 3-pt shear and convergence correlation functions which had been an important missing link between weak lensing three-point statistics. As an intermediate step, we found the 2-pt analogue of these relations and proved that it is the non-symmetrized form of the existing $\xi_+-\xi_-$ relation. By drawing analogy to the corresponding 1-pt relations, namely the Kaiser-Squires relation and its isotropic form, we have further revealed that the newly derived relations and already established results fit into the same theoretical framework. The consistency of the configuration space relations with the known Fourier space relations have also been shown.

The 3-pt relations derived are simple both analytically and numerically. They can be used as an alternative way of relating the measurable 3-pt weak lensing statistics with the statistics of the underlying matter density field. Up to now one has to use the relations between the $\gamma$3PCFs and the convergence bispectrum to link theory to the observable 3-pt shear signal. Since the $\gamma$3PCFs are very oscillatory and complicated functions of the convergence bispectrum \citep{s05}, it is hard to study the behavior of the 3-pt shear signal for a given convergence bispectrum model. With the relations we derived, one can instead study the properties of the 3-pt shear signal by constructing models for the $\kappa$3PCFs. 

The method we used to derive these relations is based on the relation between the 2-pt correlation functions of the lensing deflection potential and the convergence. The same method also allows one to systematically derive the relations between correlations functions of other weak lensing quantities, including the deflection potential $\psi$, the shear $\gamma$, the convergence $\kappa$, and the deflection angle $\alpha$. We present the forms of some 2- and 3-pt relations in Appendix.\;\ref{sec:mixcor}. Some of them are potentially of interest for studies of galaxy-galaxy(-galaxy) lensing and lensing of the Cosmic Microwave Background.

Since the relations we obtained have complex kernels with non-trivial spin number and singularities, special care is needed when they are used numerically. We demonstrated how the numerical evaluation can be done, in particular the design of the sampling grid. Examples of numerical evaluation were shown for both 2- and 3-pt relations using toy models for the convergence correlation function. Their results match very well with the analytical expectations.

Separating E- and B-modes from measurements of the $\gamma$3PCFs is particularly important since the systematic effects at the 3-pt level are less understood. So far the only 3-pt statistics allowing for an E/B-mode decomposition is the aperture mass statistics \citep{jarvis04, s05} which is plagued with the same problem as the 2-pt aperture statistics pointed out by \citet{kilbinger06}. To amend this problem, one needs to construct the 3-pt correspondence of the newly developed 2-pt statistics \citep{SK07,eifler10,fu10,s10} which allows for an E/B-mode decomposition on a finite interval. As a direct theoretical application of the 3-pt relations derived in this paper, we used them to formulate the conditions for E/B-mode decomposition of lensing 3-pt statistics, in analogy to the 2-pt condition given by \citet{SK07}. These conditions are the basis of formulating additional constraints which lead to E/B-mode decomposition over a finite region, therefore they provide a starting point for future works on constructing better 3-pt shear statistics. 

Our work was done for the case of weak lensing, but since it has only used the mathematical structure of the shear and the convergence, it applies also to other 2D polarization fields such as the polarization fluctuations of the Cosmic Microwave Background.

\begin{acknowledgements}
XS would like to thank Fujun Du, Guangxing Li, and Andy Taylor for helpful discussions. BJ acknowledges support by a UK Space Agency Euclid grant. This work was supported by the Deutsche Forschungsgemeinschaft under the Collaborative Research Center TR-33 `The Dark Universe' and the EC within the RTN Network `DUEL'.

\end{acknowledgements}

\begin{appendix}

\section{Fourier transform of the $F$ and $G_0$ kernels}
\label{app:1}
The $F$ and $G_0$ kernels, as defined in (\ref{eq:defineF}) and (\ref{eq:defineG}) relate the 2- and 3-pt shear and convergence correlation functions, respectively. In Sect.\;$2$ we have derived their explicit forms, see (\ref{eq:Fform}) and (\ref{eq:G2p}). According to the Fourier space relations of the shear and convergence statistics (\ref{eq:FourierF}) and (\ref{eq:FourierG}), one expects that $F/\pi^2$ and $\expo{4\ic\beta}$ are Fourier pairs, as well as $-G_0/\pi^3$ and $\expo{2\ic\br{\beta_1+\beta_2+\beta_3}}$. Here we perform the Fourier transforms of $F/\pi^2$ and $-G_0/\pi^3$, with $F$ and $G_0$ given in (\ref{eq:Fform}) and (\ref{eq:G2p}).

The Fourier transformation of the kernel $F$ is a 2D integral of the form
\eq{
\label{eq:FFtransf}
\frac{\tilde{F}(\vek{\ell})}{\pi^2} = \int \dd^2 a \; \expo{-\ic \svek{\ell}\cdot\svek{a}} \;\frac{F(\vek{a})}{\pi^2} =  \frac{2}{\pi} \int \dd^2 a \; \expo{-\ic \svek{\ell}\cdot\svek{a}}\;\frac{\vek{a}}{\vek{a}^{*3}}\,.
}
The Fourier transformation of the kernel $G_0$ can be greatly simplified by performing translational shifts to the integration variables. It turns out that the full transformation is composed of 2D integrals similar to that in (\ref{eq:FFtransf}), 
\eqs{
-\frac{\tilde{G}_0(\vek{\ell_1},\vek{\ell_2})}{\pi^3}= &\int \dd^2 a \;\int \dd^2 b\; \expo{-\ic \br{\svek{\ell_1}\cdot\svek{a}+\svek{\ell_2}\cdot\svek{b}}} \;\br{-\frac{G_0(\vek{a},\vek{b})}{\pi^3}}\\
=& -\frac{2}{\pi^2} \int \dd^2 a  \expo{-\ic \svek{\ell_1}\cdot\svek{a}} \;\int \dd^2 b  \expo{-\ic \svek{\ell_2}\cdot\svek{b}} \; \bb{\frac{1}{(\vek{a}^*-\vek{b}^*)^2} \br{\frac{\vek{a}}{\vek{a}^{*3}}+\frac{\vek{b}}{\vek{b}^{*3}}} + \frac{1}{(\vek{a}^*-\vek{b}^*)^3}\br{\frac{\vek{a}}{\vek{a}^{*2}}-\frac{\vek{b}}{\vek{b}^{*2}}} }\\
=& -\frac{2}{\pi^2} \int \dd^2 a  \expo{-\ic \svek{\ell_1}\cdot\svek{a}}  \frac{\vek{a}}{\vek{a}^{*3}} \; \int \dd^2 b  \expo{-\ic \svek{\ell_2}\cdot\svek{b}}  \frac{1}{(\vek{a}^*-\vek{b}^*)^2} 
-\frac{2}{\pi^2} \int \dd^2 b  \expo{-\ic \svek{\ell_2}\cdot\svek{b}} \frac{\vek{b}}{\vek{b}^{*3}}\; \int \dd^2 a  \expo{-\ic \svek{\ell_1}\cdot\svek{a}}  \frac{1}{(\vek{a}^*-\vek{b}^*)^2}  \\
& -  \frac{2}{\pi^2} \int \dd^2 a  \expo{-\ic \svek{\ell_1}\cdot\svek{a}}  \frac{\vek{a}}{\vek{a}^{*2}} \; \int \dd^2 b  \expo{-\ic \svek{\ell_2}\cdot\svek{b}}  \frac{1}{(\vek{a}^*-\vek{b}^*)^3}
+ \frac{2}{\pi^2} \int \dd^2 b  \expo{-\ic \svek{\ell_2}\cdot\svek{b}} \frac{\vek{b}}{\vek{b}^{*2}}\; \int \dd^2 a  \expo{-\ic \svek{\ell_1}\cdot\svek{a}}  \frac{1}{(\vek{a}^*-\vek{b}^*)^3}\\
=& -\frac{2}{\pi^2} \int \dd^2 a  \expo{-\ic (\svek{\ell_1}+\svek{\ell_2})\cdot\svek{a}}  \frac{\vek{a}}{\vek{a}^{*3}} \; \int \dd^2 b  \expo{-\ic \svek{\ell_2}\cdot\svek{b}}  \frac{1}{\vek{b}^{*2}} 
-\frac{2}{\pi^2} \int \dd^2 a  \expo{-\ic (\svek{\ell_1}+\svek{\ell_2})\cdot\svek{a}}  \frac{\vek{a}}{\vek{a}^{*3}} \; \int \dd^2 b  \expo{-\ic \svek{\ell_1}\cdot\svek{b}}  \frac{1}{\vek{b}^{*2}}   \\
& + \frac{2}{\pi^2} \int \dd^2 a  \expo{-\ic (\svek{\ell_1}+\svek{\ell_2})\cdot\svek{a}}  \frac{\vek{a}}{\vek{a}^{*2}} \; \int \dd^2 b  \expo{-\ic \svek{\ell_2}\cdot\svek{b}}  \frac{1}{\vek{b}^{*3}}
+ \frac{2}{\pi^2} \int \dd^2 a  \expo{-\ic (\svek{\ell_1}+\svek{\ell_2})\cdot\svek{a}}  \frac{\vek{a}}{\vek{a}^{*2}} \; \int \dd^2 b  \expo{-\ic \svek{\ell_1}\cdot\svek{b}}  \frac{1}{\vek{b}^{*3}}\,.
}
Performing the 2D integrals in polar coordinates results in
\eq{
\label{eq:11}
\int \dd^2 a  \expo{-\ic \svek{\ell}\cdot\svek{a}}  \frac{\vek{a}}{\vek{a}^{*3}} = \int_0^{\infty} \frac{\dd a}{a} \; \int_0^{2\pi} \dd \phi_a
 \expo{4\ic\phi_a} \expo{-\ic \ell a \cos(\phi_a-\beta)} = 2\pi \expo{4\ic\beta} \int_0^{\infty} \frac{\dd a}{a} J_4(\ell a) = \frac{\pi}{2} \expo{4\ic\beta} \,,
}

\eq{
\int \dd^2 a  \expo{-\ic \svek{\ell}\cdot\svek{a}}  \frac{\vek{a}}{\vek{a}^{*2}} = \int_0^{\infty} \dd a \; \int_0^{2\pi} \dd \phi_a
 \expo{3\ic\phi_a} \expo{-\ic \ell a \cos(\phi_a-\beta)} = 2\pi\ic \expo{3\ic\beta} \int_0^{\infty} \dd a \; J_3(\ell a) = 2\pi\ic \frac{\expo{3\ic\beta} }{\ell}\,,
}

\eq{
\int \dd^2 a  \expo{-\ic \svek{\ell}\cdot\svek{a}}  \frac{1}{\vek{a}^{*3}} = \int_0^{\infty} \frac{\dd a}{a^2} \; \int_0^{2\pi} \dd \phi_a
 \expo{3\ic\phi_a} \expo{-\ic \ell a \cos(\phi_a-\beta)} = 2\pi \ic \expo{3\ic\beta} \int_0^{\infty} \frac{\dd a}{a^2} J_3(\ell a) = \frac{\ic \pi\ell}{4} \expo{3\ic\beta} \,,
}
and
\eq{
\int \dd^2 a  \expo{-\ic \svek{\ell}\cdot\svek{a}}  \frac{1}{\vek{a}^{*2}} = \int_0^{\infty} \frac{\dd a}{a} \; \int_0^{2\pi} \dd \phi_a
 \expo{2\ic\phi_a} \expo{-\ic \ell a \cos(\phi_a-\beta)} = - 2\pi \expo{2\ic\beta} \int_0^{\infty} \frac{\dd a}{a} J_2(\ell a) = - \pi \expo{2\ic\beta} \,.
}

Combining (\ref{eq:FFtransf}) and (\ref{eq:11}) yields 
\eq{
\label{eq:FFtransf1}
\frac{\tilde{F}(\vek{\ell})}{\pi^2} =  \frac{2}{\pi}\br{ \frac{\pi}{2} \expo{4\ic\beta} }= \expo{4\ic\beta}\,,
}
which demonstrates that the Fourier transformation of $F/\pi^2$ is indeed the phase factor $\expo{4\ic\beta}$ in (\ref{eq:FourierF}).

For the $G_0$ kernel we still need to take account that $\vek{\ell_3} = \ell_3 \expo{\ic \beta_3} \equiv -\vek{\ell_1}-\vek{\ell_2}$, so that
\eqs{
-\frac{\tilde{G}_0(\vek{\ell_1},\vek{\ell_2})}{\pi^3} =\;& -\frac{2}{\pi^2} \br{\frac{\pi}{2} \expo{4\ic\beta_3} } \br{- \pi \expo{2\ic\beta_2} }  
   - \frac{2}{\pi^2}  \br{2\pi\ic \frac{\expo{3\ic\beta_3} }{\ell_3} }  \br{\frac{\ic \pi\ell_2}{4} \expo{3\ic\beta_2} }    
   -\frac{2}{\pi^2} \br{\frac{\pi}{2} \expo{4\ic\beta_3} } \br{- \pi \expo{2\ic\beta_1} }  
   - \frac{2}{\pi^2}  \br{2\pi\ic \frac{\expo{3\ic\beta_3} }{\ell_3} }  \br{\frac{\ic \pi\ell_1}{4} \expo{3\ic\beta_1} }\\
=\;& \expo{4\ic\beta_3 + 2\ic\beta_2 }   + \frac{\ell_2}{\ell_3} \expo{3\ic\beta_3+3\ic\beta_2}  +\expo{2\ic\beta_1 + 4\ic\beta_3 } + \frac{\ell_1}{\ell_3} \expo{3\ic\beta_1+3\ic\beta_3} =\;\frac{\ell_3^2}{\ell_3^{*2}}\frac{\ell_2}{\ell_2^*} + \frac{\ell_2^2}{\ell_2^*}\frac{\ell_3}{\ell_3^{*2}} + \frac{\ell_3^2}{\ell_3^{*2}}\frac{\ell_1}{\ell_1^*} + \frac{\ell_1^2}{\ell_1^*}\frac{\ell_3}{\ell_3^{*2}}\\ 
=\;& \frac{\ell_1\ell_3\br{\ell_1+\ell_3}}{\ell_1^*\ell_3^*}+\frac{\ell_2\ell_3\br{\ell_2+\ell_3}}{\ell_2^*\ell_3^*} = -\frac{\ell_1\ell_2\ell_3}{\ell_3^{*2}} \br{\frac{1}{\ell_1^*}+\frac{1}{\ell_2^*}} = \frac{\ell_1\ell_2\ell_3}{\ell_1^*\ell_2^*\ell_3^*} = \expo{2\ic\br{\beta_1+\beta_2+\beta_3}}\,.
}
Thus the Fourier transformation of $-G_0/\pi^3$ equals the phase factor $\expo{2\ic\br{\beta_1+\beta_2+\beta_3}}$ in (\ref{eq:FourierG}), as expected.

\section{Relations between other correlation functions}
\label{sec:mixcor}

In Sect.\;\ref{sec:formsofk} we have derived the relations between the shear and the convergence 2- and 3-pt correlation functions. The method we used is based on the relation between the 2-pt correlation functions of the convergence $\kappa$ and the deflection potential $\psi$ (\ref{eq:2kpsi2}). Thus the method can easily be generated to derive relations between the correlation function of $\kappa$ and that of any weak lensing quantity $g$ which can be expressed as derivatives of $\psi$. We denote $g = \rm{D}_g \psi$, and write these 2-pt relations in a general form  
\eq{
\label{eq:2pgen}
\ba{g(\vek{x_1}) g'(\vek{x_2})} = \frac{1}{\pi^2} \int \dd^2 y \; \ba{\kappa\kappa}(\vek{y}) \; \mathcal{H}(\vek{x_1}-\vek{x_2}-\vek{y}) \,.
}
\noindent Listed below are  some candidates for $g$ and the corresponding operator $\rm{D}_g$, where $\alpha$ is the deflection angle defined as $\alpha = \partial \psi$. 
\begin{center}
\begin{tabular}[h]{c||c|c|c|c}
$g$           &  $\psi$ & $\kappa$ & $\gamma$ & $\alpha$ \\
\hline
$\rm{D}_g$   & 1 & $\nabla^2/2$  & $\partial^2/2$ & $\partial$ \\
\end{tabular}
\end{center}

Let $\rm{D}_g \rm{D}_{g'}$ act on both sides of the relation between $\ba{\psi\psi}$ and $\ba{\kappa\kappa}$ (\ref{eq:2kpsi}), and use the statistical homogeneity of the $\kappa$ field, one can obtain an integration form of the kernel $\mathcal{H}$, in analogy to (\ref{eq:fkerneldef}). Let the same operator act on both sides of (\ref{eq:2kpsi2}), one can express $\mathcal{H}$ as derivatives of the convolution kernel $\cal{F}$ in (\ref{eq:2kpsi2}), as $\mathcal{H}(\vek{x_1}-\vek{x_2}-\vek{y}) = \rm{D}_g \rm{D}_{g'} \cal{F}(\vek{x_1}-\vek{x_2}-\vek{y})$. Further inserting the explicit form of $\cal{F}$ (\ref{eq:Fform}) allows one to obtain the explicit form of $\mathcal{H}$ as a function of $\vek{z} = \vek{x_1}-\vek{x_2}-\vek{y}$. We summarize some of the 2-pt relations in Table.\;\ref{tab:2pre}. Note that the form of $\mathcal{H}(\cal{F})$ for $\ba{\alpha\alpha}$ has a minus sign, which is due to the fact that $\partial_{x_1}\partial_{x_2}\cal{F}(\vek{x_1}-\vek{x_2}-\vek{y})$ $= - {\partial^{2}}  \cal{F}(\vek{z})$ with $\vek{z} = \vek{x_1}-\vek{x_2}-\vek{y}$.

\begin{table*}
\centering
\caption{Forms of the convolution kernel $\mathcal{H}$ as defined in (\ref{eq:2pgen}) for different weak lensing 2-pt statistics.} 
\label{tab:2pre}
\begin{tabular}[h]{c|c|c|c}
 $\ba{XX}$ & $\mathcal{H}(\cal{F})$ &   integration form of $\mathcal{H}$ & $\mathcal{H}(\vek{z})$\\
\hline\hline  

$\ba{\psi\psi}$     &   $\cal{F}$              &   $\int \dd^2 v \; \ln|\vv| \ln|\vz-\vv|$   &   $\br{\pi/2} |\vz|^2 \br{\ln{\vz}-1} $ \\
$\ba{\gamma\psi}$ &  $\frac{1}{2}\partial^2\cal{F}$ & $- \int \dd^2 v \; \frac{1}{(\vv^*-\vzs)^2} \ln|\vv|$  &  $\br{\pi/2} \vz/\vzs $\\
$\ba{\alpha\alpha}$  &   $-\partial^2\cal{F}$  &  $\int \dd^2 v \; \frac{1}{\vek{v}^{*}} \frac{1}{\vv^*-\vzs}$  &   $-{\pi} \vz/\vzs$\\
$\ba{\kappa\psi}$  &   $\frac{1}{2}\partial\partial^*\cal{F}$   &    $\pi\int \dd^2 v \;\delta^{(2)}_{\rm D}(\vz-\vv) \ln|\vv|$  &  $\pi\ln{\vz}$\\
$\ba{\alpha\gamma}$   &   $\frac{1}{2}\partial^3\cal{F}$   &  $\int \dd^2 v \; \frac{1}{\vek{v}^{*2}}\frac{1}{\vv^*-\vzs}$   &  $-{\pi} \vz/\vz^{*2}$\\
$\ba{\alpha\kappa}$   &   $\frac{1}{2}\partial^2\partial^*\cal{F}$   &  $\pi\int \dd^2 v \; \frac{1}{\vzs-\vv^*}\delta^{(2)}_{\rm D}(\vv)$   &  $\pi/\vzs$\\
$\ba{\alpha^*\gamma}$   &   $\frac{1}{2}\partial^2\partial^*\cal{F}$ &  $ \int \dd^2 v \; \frac{1}{\vek{v}^{*2}}\frac{1}{\vv-\vz}$   &  $\pi/\vzs$\\
$\ba{\gamma\gamma}$  &   $\frac{1}{4}\partial^4\cal{F}$  &  $\int \dd^2 v \; \frac{1}{\vek{v}^{*2}} \frac{1}{(\vv^*-\vzs)^2}$  &   $2\pi \vz/\vz^{*3}$\\
$\ba{\kappa\gamma}$   &   $\frac{1}{4}\partial^3\partial^*\cal{F}$   &  $-\pi\int \dd^2 v \; \frac{1}{\vek{v}^{*2}}\delta^{(2)}_{\rm D}(\vz-\vv)$   &  $-\pi/\vz^{*2}$\\
$\ba{\kappa\kappa}$   &   $\frac{1}{4}\partial^2\partial^{*2}\cal{F}$   &      &  $\pi^2\delta^{(2)}_{\rm D}(\vz)$\\
$\ba{\gamma\gamma^*}$   &   $\frac{1}{4}\partial^2\partial^{*2}\cal{F}$   & $\int \dd^2 v \; \frac{1}{\vek{v}^{2}} \frac{1}{(\vv^*-\vzs)^2}$      &  $\pi^2\delta^{(2)}_{\rm D}(\vz)$\\
   
\end{tabular}
\renewcommand{\footnoterule}{}  
\end{table*}

We write the relations between $\ba{\kappa\kappa\kappa}$ and the 3-pt correlation functions of the $g$'s also in a uniform convolutional form, 
\eq{
\label{eq:3pgen}
\ba{g g' g''}(\vek{x_1},\vek{x_2}) = \frac{1}{\pi^3} \int \dd^2 y_1 \int \dd^2 y_2 \; \ba{\kappa\kappa\kappa}(\vek{y_1},\vek{y_2}) \; I(\vek{x_1}-\vek{y_1},\vek{x_2}-\vek{y_2}) \,.
}
To find the explicit form of the convolution kernel $I$, we first write it into an integral form, in analogy to (\ref{eq:gkerneldef}), and then split it into terms which can be expressed also as derivatives of the kernel $\cal{F}$, like (\ref{eq:spplit}). Then with the explicit form of $\cal{F}$ one can reach the explicit form of $I$. We list the forms of the convolution kernel $I$ for some 3-pt statistics in Table.\;\ref{tab:3pre}.

\begin{table*}
\centering
\caption{Forms of the convolution kernel $I$ as defined in (\ref{eq:3pgen}) for different weak lensing 3-pt statistics.} 
\label{tab:3pre}
\begin{tabular}[h]{c|c|c|c}
 $\ba{XXX}$ &  integration form of $I$ & split form of $I$  &  $I(\vek{a},\vek{b})$\\
\hline\hline  

$\ba{\alpha\alpha\alpha}$  &  $-\int \dd^2 v \; \frac{1}{\vvs} \frac{1}{\vvs-\vas} \frac{1}{\vvs-\vbs}$  & $-\frac{1}{\vas-\vbs}\int \dd^2 v \frac{1}{\vvs}\br{\frac{1}{\vvs-\vas}-\frac{1}{\vvs-\vbs}}$   &  $\frac{\pi}{\vas-\vbs}\br{\frac{\va}{\vas}-\frac{\vb}{\vbs}} $\\
$\ba{\alpha\alpha\gamma}$   &  $-\int \dd^2 v \; \frac{1}{\vvs^2}\frac{1}{\vvs-\vas}\frac{1}{\vvs-\vbs}$  & $-\frac{1}{\vas-\vbs}\int \dd^2 v \frac{1}{\vvs^2}\br{\frac{1}{\vvs-\vas}-\frac{1}{\vvs-\vbs}}$   &   $\frac{\pi}{\vas-\vbs}\br{\frac{\va}{\vas^2}-\frac{\vb}{\vbs^2}}$ \\
$\ba{\gamma\gamma\alpha}$   &  $-\int \dd^2 v \; \frac{1}{\vvs}\frac{1}{(\vvs-\vas)^2}\frac{1}{(\vvs-\vbs)^2}$  & $-\frac{1}{(\vas-\vbs)^2}\int \dd^2 v \frac{1}{\vvs}\br{\frac{1}{(\vvs-\vas)^2}+\frac{1}{(\vvs-\vbs)^2}} $   &   $-\frac{\pi}{(\vas-\vbs)^2} \br{\frac{\va}{\vas^2} + \frac{\vb}{\vbs^2}} - \frac{2\pi}{(\vas-\vbs)^3} \br{\frac{\va}{\vas} - \frac{\vb}{\vbs}} $ \\
  &  &  $+ \frac{2}{(\vas-\vbs)^3}\int \dd^2 v \frac{1}{\vvs}\br{\frac{1}{\vvs-\vas} - \frac{1}{\vvs-\vbs}}$ &  \\
$\ba{\gamma\gamma\kappa}$   &  $\pi \int \dd^2 v \; \delta^{(2)}_{\rm D}(\vv)\frac{1}{(\vvs-\vas)^2}\frac{1}{(\vvs-\vbs)^2}$  &  &  ${\pi}/\br{\vas^2\vbs^2} $ \\
$\ba{\gamma\kappa\kappa}$   &  $\pi^2 \int \dd^2 v \; \delta^{(2)}_{\rm D}(\vv) \;\delta^{(2)}_{\rm D}(\vv-\vb) \frac{1}{(\vvs-\vas)^2}$  &  &  ${\pi^2 \delta^{(2)}_{\rm D}(\vb) }/{\vas^2} $ \\
$\ba{\alpha\alpha\kappa}$   &  $\pi \int \dd^2 v \; \delta^{(2)}_{\rm D}(\vv)\frac{1}{\vvs-\vas}\frac{1}{\vvs-\vbs}$  &  &  ${\pi}/\br{\vas\vbs}$  \\

\end{tabular}
\renewcommand{\footnoterule}{}  
\end{table*}

Some of these relations, e.g. those for $\ba{\kappa\gamma}$, $\ba{\gamma\gamma\kappa}$, and $\ba{\gamma\kappa\kappa}$, can find their application in galaxy-galaxy(-galaxy) lensing. Some other relations, e.g. those for $\ba{\alpha\alpha}$, $\ba{\alpha\kappa}$, and $\ba{\alpha\alpha\kappa}$, are potentially of interest for studies of the lensing effects on the Cosmic Microwave Background and its cross-correlation with galaxy weak-lensing maps \citep{hu00}.

\end{appendix}

\end{document}